\documentclass[twocolumn]{aastex62}

\usepackage{nicefrac}
\usepackage{savesym}
\savesymbol{tablenum}
\usepackage{siunitx}
\usepackage{amsmath}
\usepackage{booktabs}
\usepackage{multirow}

\restoresymbol{SIX}{tablenum}
\DeclareSIUnit\parsec{pc}
\DeclareSIUnit\years{yr}
\DeclareSIUnit\solmass{M_{\odot}}
\DeclareSIUnit\sollum{L_{\odot}}
\DeclareSIUnit\AU{au}
\DeclareSIUnit\om{\Omega}
\DeclareSIUnit\orb{T_{\mathrm{orb}}}
\DeclareSIUnit\scaleheight{H}

\definecolor{dodgerblue}{rgb}{0.11764706, 0.56470588, 1.}
\definecolor{seagreen}{rgb}{0.18039216, 0.54509804, 0.34117647}
\definecolor{maroon}{rgb}{0.50196078, 0., 0.}

\submitjournal{ApJ}

\shorttitle{Stability of Protoplanetary Accretion Disks}
\shortauthors{Pfeil \& Klahr}

\newcommand{\ADD}[1]{{#1}}
\newcommand{\CORR}[1]{{#1}}

\begin{document}

\title{Mapping the Conditions for Hydrodynamic Instability on Steady State Accretion Models of Protoplanetary Disks}

\author[0000-0002-4171-7302]{Thomas Pfeil}
\affiliation{\centering Max-Planck-Institut f\"ur Astronomie, K\"onigstuhl 17, D-69117 Heidelberg, Germany}
\correspondingauthor{Thomas Pfeil \& Hubert Klahr}
\email{pfeil@mpia.de}
\email{klahr@mpia.de}

\author[0000-0002-8227-5467]{Hubert Klahr}
\affiliation{\centering Max-Planck-Institut f\"ur Astronomie, K\"onigstuhl 17, D-69117 Heidelberg, Germany}

\begin{abstract}
Hydrodynamical instabilities in disks around young stars depend on the thermodynamic stratification of the disk and on the local rate of thermal relaxation. Here, we map the spatial extent of unstable regions for the {\em Vertical Shear Instability} (VSI), the {\em Convective OverStability} (COS), and the amplification of vortices via the {\em Subcritical Baroclinic Instability} (SBI). 
We use steady state accretion disk models, including stellar irradiation, accretion heating and radiative transfer. 
We determine the local radial and vertical stratification and thermal relaxation rate in the disk, in dependence of the stellar mass, disk mass and mass accretion rate. 
We find that passive regions of disks - i.e. the midplane temperature dominated by irradiation - are COS unstable about one pressure scale height above the midplane and VSI unstable at radii $> 10 \, \text{au}$. Vortex amplification via SBI should operate in most parts of active and passive disks. For active parts of disks (midplane temperature determined by accretion power) COS can become active down to the midplane. Same is true for the VSI because of the vertically adiabatic stratification of an internally heated disk. If hydro instabilities or other non-ideal MHD processes are able to create $\alpha$-stresses ($> 10^{-5}$) and released accretion energy leads to internal heating of the disk, hydrodynamical instabilities are likely to operate in significant parts of the planet forming zones in disks around young stars, driving gas accretion and flow structure formation. Thus hydro-instabilities are viable candidates to explain the rings and vortices observed with ALMA and VLT. 
\end{abstract}


\section{Introduction} \label{sec:intro}
Angular momentum transport and the associated accretion process in protoplanetary disks are either driven by winds \citep{Wardle_1997, Pudritz_2007, Koenigl_2010, Bai_2013} or by magnetic and hydrodynamic turbulence \citep{Luest_1952, Shakura_1973, Balbus_Hawley_1991}. One of the considerable physical processes, causing outward transport of angular momentum is the Magnetorotational Instability (MRI), which requires a sufficiently ionised shear flow in addition to weak magnetic fields. This linear instability works well in accretion disk of high temperature around black holes or neutron stars. Large parts of protoplanetary disks however have low ionisation rates and gas densities outside $\sim \SI{0.3}{\AU}$, thus non-ideal MHD effects, namely resistivity and ambipolar diffusion \citep{Lesur_2014, Gressel_2015}, largely hamper the MRI and thus open a venue for hydrodynamic instabilities, as explored by \cite{Lyra_2011}.

These hydrodynamical mechanisms include the Subcritical Baroclinic Instability (SBI) \citep{Klahr_Bodenheimer_2003, Petersen1_2007, Petersen2_2007, Lesur_Papaloizou_2010}, the Convective Overstability (COS) \citep{Klahr_Hubbard_2014}, which can be interpreted to be the linear phase of the SBI-mechanism  \citep{Lyra_2014}, and the Vertical Shear Instability (VSI) \citep{Urpin_1998, Urpin_2003, Nelson_2012, Lin_2015}, which is the protoplanetary disk equivalent of the Goldreich-Schubert-Fricke Instability \citep{Goldreich_Schubert_1967, Fricke_1968} in stars. 

The radial and vertical stratification of the disk in temperature and density and the thermal relaxation timescale  decide on whether these instabilities can exist or not. For infinite cooling times the stability constraints are given by the standard Solberg-H\o iland criteria \citep{Ruediger_2002}.
\cite{Malygin_2017} have investigated detailed models of the radiative properties of a simple, non accreting powerlaw disk profile, identifying necessary conditions for the onset of instability by mapping where the infinite cooling time condition is sufficiently violated. 

In this paper, we replace the powerlaw disk models with a self-consistent 1+1D accretion disk model that allows to determine the non-trivial temperature and density stratification of the gas as a result of gas accretion \citep{Meyer_1982, Bell_1997} and stellar irradiation \citep{DAlessio_1998}.
Whereas the surface temperature of a disk can usually be nicely approximated by a power-law, the midplane temperature can have a more complicated structure with varying gradients, reflecting the local optical depth and the rate of viscous heating \citep{Bell_1997,DAlessio_1998}. \ADD{Gas accretion and the associated heating are assumed to be the result of turbulent viscosity, determined by the free disk parameter $\alpha$ \citep{Shakura_1973}, which for our model determines the amount of thermal energy that is released inside of the disk. \\ \indent
A passive disk as defined here is to have a temperature structure in the planet-forming regions dominated by stellar irradiation (which is in our models the case for $\alpha \lesssim 10^{-5}$) and an active disk to be dominated by accretion heating in the planet-forming zone (in our models for $\alpha > 10^{-5}$). \\ \indent
The goal of this paper is to learn whether A: a passive disk is able to develop hydrodynamical instabilities and B: whether the energy release from resulting mass accretion triggered by the instabilities will support or suppress the instabilities. \\ \indent
On the other hand the mass accretion rate can also be thought to be the results of either another instability associated with non-ideal MHD effects, like 
a strongly suppressed MRI (which would otherwise quench the instabilities we are aiming to investigate) or be produced by magnetically driven disk winds or Hall MHD, also leading to some heating of the disk.}

The knowledge of the physical conditions inside of a protoplanetary disk makes it possible to determine where the necessary criteria for instability are met and how fast the corresponding linear perturbations grow with time. This is crucial to understand the nature of angular momentum transport in circumstellar disks and to set-up simulations of hydrodynamic instabilities in realistically modelled physical environments.

The weak hydrodynamical instabilities are also of special interest for planet formation theory, because even if their contribution to angular momentum transport might be small, they drive the formation of non laminar flow features like zonal-flows and vortices. Such pressure maxima are able to accumulate the inwards drifting dust particles and could therefore be the birthplaces of  planetesimals and planets  \citep{Barge_1995, Klahr_2006}. 
Such structures are observed lately in circumstellar disks by ALMA and VLT \citep{Marel_2013, Carrasco_2016}.
For a recent review on the role of non-laminar flow features on planetesimal formation we refer to \cite {Klahr2018}. \ADD{\cite{Hartmann_2018} revisited the physics of the accretion process itself and found that even a relatively low $\alpha \gtrsim 10^{-4}$ is enough to explain the observed accretion rates onto T-Tauri stars. They conclude that the hydrodynamical contribution to angular momentum transport might have been underestimated in the past and that in some cases hydrodynamic turbulence might even be more important to the accretion than magnetic contributions originating from the MRI.}

\begin{figure}[ht] 
\centering
\includegraphics[width=0.455\textwidth]{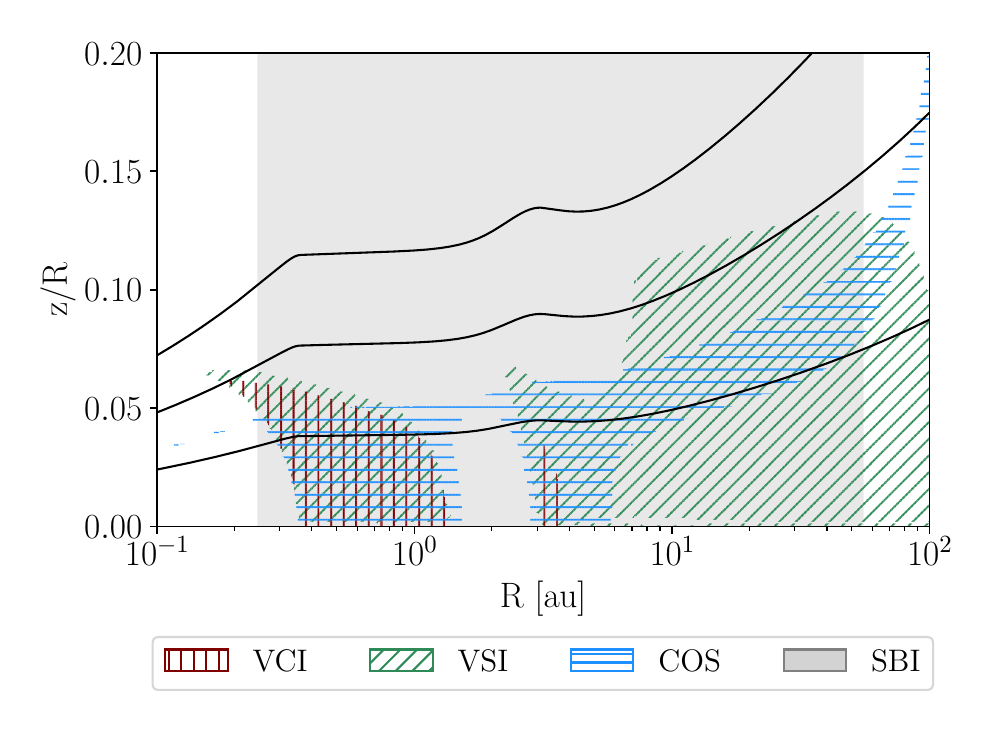}
\caption{Stability Map for a Solar Nebula like disk around a solar mass star with $M_{\mathrm{disk}}=\SI{0.1}{\solmass}$ and a moderate local viscous heating of $\alpha=10^{-3}$. The black lines indicate 1, 2 and 3 pressure scale heights respectively. {\bf \color{seagreen} Vertical Shear Instability (VSI - green lines slanted to the right)} can occur at larger radii, where cooling is efficient or alternatively in the inner more optical thick parts of the nebula, which is vertically adiabatic, indicated by the possible occurrence of {\bf \color{maroon} Vertical Convective Instability (VCI - vertical red lines)}. {\bf \color{dodgerblue} Convective Overstability (COS - horizontal blue lines)} will occur in the irradiation dominated outer parts only at a height of 1 pressure scale height and above, because the entropy gradient flips sign only some height above the midplane. In the inner parts of the nebula, where viscous heating is important and especially in the opacity transition zones (ice line \SIrange{3}{6}{\AU} and silicates evaporation zones \SIrange{0.4}{1.5}{\AU}) the entropy gradient is negative down to the midplane, forming a sweet-spot for COS. The {\color{gray}\bf Convective Amplification of large scale vortices (SBI - Grey Shaded region)} can occur through out most of the disk (\SIrange{0.3}{50}{\AU}), because the surface density structure is shallower than the midplane density, thus radial buoyancy (negative gradient of vertical integrated entropy) can much easier be established.}
\label{map}
\end{figure}

In the following section we give an overview of the basic physics of the investigated instabilities, their analytical growth rates, and the concepts of thermal relaxation used in the scope of this work. The 1+1D disk model as well as the used opacity model are described in \autoref{sec:models}.
The general influence of the disk structure on stability, as well as the spatial distributions and growth rates of the introduced mechanisms are presented in \autoref{sec:results}.
Our stability maps, shown in \autoref{subsec:stability} sum up the gained knowledge of the distribution of the susceptible regions for the investigated instabilities and  parameter sets. One example of such a map is shown in \autoref{map}, for a disk model with input parameters $M_{\mathrm{disk}}=\SI{0.1}{\solmass}$, a moderate local viscous heating of $\alpha=10^{-3}$ and a central star of $M_*=\SI{1}{\solmass}$. We finally summarise and conclude in \autoref{sec:conclusion}.

\section{Instabilities} \label{sec:instabilities}
The hydrodynamic instabilities discussed here arise either from vertical shear or from the radial buoyancy in the disk. Both are linked to the radial gradient of temperature. 

The {\it Vertical Shear Instability} is a special case for the violation of the Rayleigh Criterion, i.e.\ that gas by moving upward can move also outward under conservation of the specific angular momentum, which leads to a release of kinetic energy.
The {\it Convective Overstability} and its weakly non-linear extension the {\it Subcritical Baroclinic Instability} are both special cases of thermal convection in the radial direction of the disk, i.e. of a radial super-adiabatic stratification. 

\ADD{Another hydrodynamical effect creating vortices in disks is the so called {\it Zombie Vortex Instability} \citep{Marcus_2014, 2016ApJ...830...95U}, but we will not include it in our investigation, because 
it is a non-linear instability, making predictions on its occurence more complicated than the for the linear instabilities we study here.
A study by \cite{Lesur_Latter_2016} suggests that the instability may occur in the inner optical thick regions of disks ($R < \SI{0.3}{\AU} $), where thermal relaxation takes long enough to allow for the necessary vertical internal gravity waves for the ZVI to operate. On the contrary \cite{2018arXiv181006588B} argue that at larger distances from the star ($R > \SI{2}{\AU}$ or $R > \SI{16}{\AU}$ dependent on the assumptions of dust growth and settling) at sufficient height above the midplane ($z > \SI{1.5}{\scaleheight}$), dust might thermally decouple from the gas, allowing the ZVI to operate, provided gas opacities to be small (see the discussion in \citet{Malygin_2017}). \cite{2018arXiv181006588B} also find in numerical experiments that even in cases where ZVI operates only above $z > \SI{1.5}{\scaleheight}$ still some turbulence reaches the midplane.
In summary the conditions for ZVI are currently more a problem of the proper dust size and gas opacity model, rather than the influence of accretion onto the temperature gradients in the disk, which is the actual topic of our paper. We therefore abstain from mapping out likely ZVI regions, while noting that the dust size distributions for protoplanetary disks constrained by observations combined with numerical models of the growth-processes (e.g. \cite{Birnstiel2012, Estrada2016}) will hopefully bring light into these outstanding issues. 
%
%
}

The following sections briefly review the mechanisms for which we have predictions of linear growth-rates. \ADD{We want to point out that the growth rates 
presented in the following sections are derived for disk models without vertical temperature stratification.
The influences of disk structure and viscosity on the instabilities are further discussed in the following, while the treatment of radial flows was not incorporated in this study. \\ \indent 
Our models do not include the effects of magnetic fields, we are thus not able to treat their impacts on the evolution of the hydrodynamical instabilities, beyond providing some heating of the disk as incorporated by $\alpha$.
We explicitly assume that the non-ideal MHD terms do
allow for sufficient diffusion of magnetic fields thus hydro-dynamical instabilities will not be suppressed 
\citep{Lyra_2011,Latter_2018}.}

\subsection{Convective Overstability}\label{subsec:cos}
\citet{Klahr_Hubbard_2014} considered finite thermal relaxation times ($\tau_{\mathrm{relax}}$) in their linear, inelastic stability analysis of protoplanetary disks and found a new, thermally driven instability, which can be described as radial convection on epicycles. The mechanisms can be explained as follows. An outward perturbation of a fluid element brings it into contact with a cooler surrounding, due to the radial temperature gradient of the disk. While the gas parcel undergoes half an epicycle it changes its temperature due to heat exchange with its surrounding on a timescale $\tau_{\mathrm{relax}}$. When it arrives at the initial radial distance to the star, its entropy is lower than the initial value, which leads to an inward acceleration due to buoyancy. On its extended half epicycle through the inner and hotter region of the disk, it undergoes an increase of temperature and entropy. When it finally arrives at the initial radius, it experiences an outward buoyancy force. This positive feedback, caused by a radially buoyant stratification is called Convective Overstability (COS). The linear phase of the COS drives motions in the disk's $R-z$ plane with a growth rate of  
\begin{equation}\label{COS_rate}
\Gamma_{\mathrm{COS}}=\frac{1}{2}\frac{-\gamma \tau_{\mathrm{relax}} N_R^2}{1+\gamma^2 \tau_{\mathrm{relax}}^2(\kappa_R^2+N_R^2)}
\end{equation}
\citep{Klahr_Hubbard_2014}, where $\gamma$ is the heat capacity ratio of the gas, $\kappa_R$ is the radial epicyclic frequency and $N_R^2$ refers to the square of the radial Brunt-V\"ais\"al\"a-frequency, which is indicating stability when positive, and buoyancy driven instability when negative. \ADD{The necessary condition for COS to develop is thus given by a radially buoyant stratification ($N_R^2 <0$), which is expected to be present in the optically thick regions of the disk. The growth rate's dependency on perturbation wavenumber therefore enters \autoref{COS_rate} via the thermal relaxation time $\tau_{\mathrm{relax}}=\tau_{\mathrm{diff}}\propto k^{-2}$ (see \autoref{subsec:therm}). The assumption of $\tau_{\mathrm{relax}}=\tau_{\mathrm{diff}}$ is justified in this case, because the necessary condition $N_R <0$ requires the disk to be optically thick. \\ \indent
The wavenumbers leading to significant growth of perturbations have to fulfil $k_z \gg k_R$, which implies radially elongated and vertically unextended motions ($k\approx k_z$) \citep{Klahr_Hubbard_2014,Lyra_2014} and thus we estimate the influence of vertical stratification onto the growth rates as unimportant. Therefore it is justified to use the \cite{Klahr_Hubbard_2014} growth rate, which was derived in a radially, but not vertically stratified disk-setup.}
The sign of $N_R^2$ directly depends on the radial (cylindrical) density and temperature structure of the disks in terms of 
\begin{eqnarray}
\beta_{T}&=\frac{\mathrm{d}\log ({T})}{\mathrm{d} \log (R)}\\
\beta_{\rho}&=\frac{\mathrm{d}\log ({\rho})}{\mathrm{d} \log (R)},
\end{eqnarray}
from which we derive the radial slope in pressure \CORR{$\beta_P = \beta_{\rho} + \beta_{T}$ and specific entropy $\beta_S = \beta_{T} + (1-\gamma)\beta_{\rho}$ }resulting in the expression 
\begin{equation}
N_R^2= - \frac{1}{\gamma}\left(\frac{H}{R}\right)^2 \beta_{{S}} \beta_{{P}} \Omega^2,
\label{NR2}
\end{equation}
\citep{2013ApJ...765..115R} where $H/R$ represents the disk's aspect ratio with respect to the local pressure scale height $H$, i.e. an expression for the temperature of the disk \ADD{and $\Omega$ is the local Keplerian angular frequency}.
Since this overstability relies on entropy differences between perturbed fluid parcels and their surrounding, relaxation times have to be neither too small nor too large. In the first case, a fluid element would always adopt the temperature of its surrounding. Its movement would be isothermal and no buoyant force would act on it, which means that Rayleigh's stability criterion applies to it. The latter case describes an adiabatic perturbation, where the fluid's entropy stays constant during its epicyclic motion. This means that it follows a stable, buoyancy adjusted epicycle \citep{Latter_2016}. 
The relaxation time for maximum growth of the linear phase was also calculated by \cite{Klahr_Hubbard_2014} to be
\begin{equation}\label{COS_max}
\tau_{\mathrm{max, COS}} =\frac{1}{\gamma \Omega}.
\end{equation} 
\ADD{This condition can be used to derive the instability's maximum growing wavenumber $k_{\mathrm{max, COS}}=1/\sqrt{\tilde{D}_{\mathrm{E}} \tau_{\mathrm{max, COS}}}$ (see \autoref{subsec:therm} \autoref{t_diff} for more detail). In this study, we are interested in the modes which fulfil this condition and thus grow fastest with a rate derived by \cite{Klahr_Hubbard_2014}
\begin{equation}\label{COSmax}
\Gamma_{\mathrm{max,COS}}=-\frac{N_R^2}{4 \Omega}.
\end{equation}
Viscosity was shown to hinder the growth of COS especially for small scale perturbations by \cite{Klahr_Hubbard_2014} and \cite{Latter_2016}, yet realistic molecular viscosity is too low
to provide an obstacle to the COS, thus these considerations are not incorporated into our studies.}
The finite amplitude perturbations created by the COS can trigger its non-linear phase, the Subcritical Baroclinic Instability (SBI) \citep{Klahr_Bodenheimer_2003, Petersen1_2007, Petersen2_2007, Lyra_2014}, which amplifies existing vortices in the disk's $R-\phi$ plane (see \autoref{subsec:sbi}). These vortices are of interest for the growth of dust to planetesimals, because they are able to accumulate dust particles \citep{Barge_1995}.

\subsection{Subcritical Baroclinic Instability}\label{subsec:sbi}
Large anticyclonic vortices are quasi 2D structures in the $R-\phi$ plane of the disk. They have vertically little variation over more than a pressure scale height above the midplane \citep{Meheut_2012, Manger_2018}. Thus to study their possible amplification in the SBI mechanism, which relies on the radial buoyancy, one has to consider their vertically integrated entropy and pressure structure. 
We therefore use the definition of a vertically integrated density \ADD{$\Sigma$} and entropy \ADD{$\tilde{S}$}, also used in \cite{Klahr_2004} and \cite{Klahr_2013}
\begin{align}
\Sigma &=\int_0^{z_{\mathrm{max}}} \rho(R,z)\mathrm{d}z \\
\tilde{S}&=C_{\mathrm{V}} \log(\tilde{K}),
\end{align}
where we assume a polytropic equation of state for a two dimensional pressure $\tilde{P} \CORR{=} \tilde{K}\Sigma^{\tilde{\gamma}}$ with an entropy-like potential temperature $\tilde{K}=\tilde{P}\Sigma^{-\tilde{\gamma}}$. The heat capacity ratio has to be adjusted to a vertically integrated value $\tilde{\gamma}$, defined by \cite{Goldreich_1986} as $\tilde{\gamma}=(3\gamma -1)/(\gamma+1)=1.354$ for a typical mixture of hydrogen and helium gas ($\gamma=1.43$). 
\ADD{The investigation of the vertically integrated disk structure has also proven to be a useful tool in the study of planet disk interactions, where the radial entropy structure impacts onto the so called horseshoe drag exerted onto a planet \citep{Baruteau_2014}. This work already showed that the radial entropy structure of the disk can be a non-monotonic function with varying and sign changing gradients.}
We now adopt the definitions of the logarithmic gradients in column density \ADD{$\beta_{\Sigma}$}, vertically integrated pressure \ADD{$\beta_{\tilde{P}}$} and entropy \ADD{$\beta_{\tilde{S}}$} from \cite{Klahr_2004}, given by
\begin{align}
\beta_{\Sigma}&=\frac{\mathrm{d}\log (\Sigma)}{\mathrm{d} \log (R)} \\
\beta_{\tilde{P}}&=\frac{\mathrm{d}\log (\tilde{K}\Sigma^{\tilde{\gamma}})}{\mathrm{d} \log (R)} \\
\beta_{\tilde{S}}&=\frac{\mathrm{d}\log (\tilde{P}\Sigma^{-\tilde{\gamma}})}{\mathrm{d} \log (R)}=\beta_{\tilde{P}}-\tilde{\gamma}\beta_{\Sigma}.
\end{align}
Using this, we find a vertically integrated, radial Brunt-V\"ais\"al\"a-frequency
\begin{equation}
\tilde{N}_R^2= - \frac{1}{\tilde{\gamma}}\left(\frac{H}{R}\right)^2 \beta_{\tilde{S}} \beta_{\tilde{P}} \Omega^2,
\end{equation}
which is formally similar to Eq.\ \ref{NR2}, but can obtain significantly different values for the same radial stratification.

The growth rate for the SBI initially suggested by \cite{Lesur_Papaloizou_2010} and modified by \cite{Beutel_2012} using a COS like stability analysis but \CORR{\cite{Goodman_1987} (GNG)} like vortices instead of plane waves is approximately
\begin{equation}
\Gamma_{\mathrm{SBI}}\approx -\frac{4 \tilde{N}_R^2}{\omega (1+\chi^2)} \left(\frac{\tilde{\gamma} \omega \tau_{\mathrm{relax}}}{1+(\tilde{\gamma}\omega \tau_{\mathrm{relax}})^2}\right).
\label{SBI_growth}
\end{equation}

In order to get an estimate for a vortex' maximum growth rate, we set $\omega \tau_{\mathrm{relax}}=1$ and determine the internal vortex angular frequency $\omega$ with the relation by GNG 
\begin{equation}
\omega=\Omega \sqrt{\frac{3}{\chi^2 - 1}}.
\end{equation}
We choose the aspect ratio of the vortex (azimuthal vs. radial extent) to be $\chi=4$, which is A: a reasonable value for vortices not be too affected by epicyclic instability \citep{2009A&A...498....1L} and B: typical for large scale 3D $R-\phi$ vortices found in numerical simulations \citep{Manger_2018}.
Combining these assumptions with \autoref{SBI_growth}, leads to an approximate growth rate of
\begin{equation}
\Gamma_{\mathrm{SBI}}\approx -\frac{\tilde{N}_R^2}{4 \Omega},
\end{equation}
which is formally the same maximum growth rate as for the COS \ADD{(\autoref{COSmax})}, but note that the vertically ``integrated'' $\tilde{N}_R^2$ can significantly differ from the height dependent $N_R(z)^2$. \ADD{This means, the local appearance of COS in a disk, which depends on $N_R(z)^2<0$ is less wide spread than SBI. But if COS occurs it can trigger the SBI, when $\tilde{N}_R^2<0$ is additionally fulfilled. \cite{2013ApJ...765..115R} confirmed the $\Gamma_{\mathrm{SBI}}\propto -\tilde{N}_R^2$ dependency in numerical experiments.}

\cite{Andrews2010} find a mean value for the radial slope of surface density in disks around young stars of $\beta_{\Sigma} = -0.9$. At the same time they estimate a temperature profile of $\beta_T=-0.6$, which leads to a radially buoyant (unstable) situation with \CORR{$\tilde{S} \propto R^{- 0.28}$, whereas for the same radial stratification the midplane density slope is $\beta_{\rho} = -2.1$ leading to a radial entropy slope of $S(z=0) \propto R^{0.3}$, which is stable.
But note that the radial density profile $\beta_{\rho}$ is a function of height $\beta_{\rho}(z) = \beta_{\rho}(z=0) + (3 + \beta_T(z)) \frac{z^2}{2H^2}$, thus in the given example starting from a height of $z = \SI{1}{H}$ and assuming a vertically constant radial temperature gradient (which is typical for the regions at large distance to the central star), one finds $\beta_{\rho}(z=\SI{1}{\scaleheight})=-0.9$. This leads to a radially declining entropy profile at one pressure scale height above the midplane with $S(z=\SI{1}{\scaleheight})\propto R^{-0.21}$}
which is again unstable, a behaviour that we will discuss in a later section, when we come to our models (see also the discussion in \citet{Umurhan2018}). 

\subsection{Vertical Shear Instability}\label{subsec:vsi}
The Vertical Shear Instability (VSI) is the analogue of the well studied Goldreich-Schubert-Fricke Instability \citep{Goldreich_Schubert_1967, Fricke_1968} for protoplanetary disks. The VSI can develop if the disk has a vertical gradient in angular frequency, \ADD{as well as the ability to cool sufficiently fast, in order to allow for vertical perturbations to develop albeit counteracting buoyancy. The vertical shear is thus a necessary condition for the onset of VSI. It can be derived from radial hydrostatic equilibrium and depends on the vertical and radial stratification of the disk
\begin{align}
\Omega(R,z)^2&=\frac{G M_*}{(R^2+z^2)^{\nicefrac{3}{2}}}+\frac{1}{R \rho}\frac{\partial P(R,z)}{\partial R} \\
\frac{\partial (\Omega R)}{\partial z}&=\frac{\partial v_{\phi}}{\partial z}=\frac{1}{2 \Omega \rho^2}\left(\frac{\partial P}{\partial z}\frac{\partial \rho}{\partial R}-\frac{\partial P}{\partial R}\frac{\partial \rho}{\partial z}\right),
\end{align}
where $v_{\phi}$ is the azimuthal flow velocity of the gas, $P$ is the pressure and, $\rho$ gives the density.}
Vertically perturbed fluid parcels, which move along the curved iso-surfaces of angular momentum, gain kinetic energy and circumvent Rayleigh's stability criterion \citep{Urpin_1998}. This instability therefore drives modes, which are vertically elongated ($k_R/k_z \gg 1$) \citep{Urpin_2004}.
\citet{Nelson_2012}, who were the first to show that the VSI can operate in protoplanetary disks, calculated the corresponding growth rate for a locally isothermal, compressible gas as
under the shearing sheet approximation  
\begin{equation}\label{VSI_growth}
\Gamma_{\mathrm{VSI}}^2=\frac{-\kappa_R^2(c_s^2 k_z^2+N_z^2)+2\Omega c_s^2 k_R k_z \frac{\partial v_{\phi}}{\partial z}}{c_s^2(k_z^2+k_R^2)+\kappa_R^2+N_z^2},
\end{equation}
where $N_z$ is the vertical buoyancy frequency, $c_s$ is the local sound speed, and $\Omega$ is the Keplerian angular frequency. They performed numerical simulation of the instability and found narrow, almost vertical motions, which caused $\alpha \sim 10^{-3}$. \ADD{Other authors, like \cite{Stoll_2014} and \cite{Manger_2018}, find turbulence associated with relatively low $\alpha$ values in the order of $\sim 10^{-4}$.}

In this work, we are interested in the maximal growing perturbation of a certain radial wavenumber. Therefore, we use the following condition by \citet{Urpin_2004} to get the corresponding fastest growing vertical wavenumber (see also \cite{2016A&A...586A..33U})
\begin{equation}\label{VSI_max}
k_z=\frac{k_R
}{2}\frac{\partial_z j^2(R,z)}{\partial_R j^2(R,z)},
\end{equation}
where $j(R,z)$ represents the specific angular momentum.
The vertically perturbed fluid parcels which are prone to be unstable to the VSI experience buoyant forces in the case of a stable stratification, which impede the instability's growth. Fast thermal relaxation can overcome this obstacle, because it adjusts the fluid parcels temperature to the background temperature and therefore diminishes buoyancy driving entropy differences.
\citet{Lin_2015} considered this effect and derived a critically slow relaxation time scale \ADD{of $\tau_{\mathrm{crit}}={\left|\partial_z v_{\phi}\right|}/{N_z^2}$} for which the VSI can grow in a convectively stable disk.
In contrast, a convectively unstable or neutral vertical stratification does not impede vertical perturbations and thus allows for VSI in the presence of sufficiently fast cooling. The behaviour of VSI in such a polytropically stratified disk was also studied by \cite{Nelson_2012} in numerical experiments,\CORR{ who found a critical relaxation time of $\tau_{\mathrm{relax}}\Omega \sim 10$. This leads to a necessary criterion for the onset of VSI in the convectively stable ($N_z^2 >0$) and unstable/neutral ($N_z^2 \leq 0$) case.
\begin{equation}
\tau_{\mathrm{crit}}=  \begin{cases}
    \frac{\left|\partial_z v_{\phi}\right|}{N_z^2}\label{VSI_crit} & \text{for  $N_z^2 >0$} \\
    \frac{10}{\Omega} & \text{for  $N_z^2 \leq 0$}
  \end{cases} 
\end{equation}
It is therefore necessary to investigate whether the vertical stratification is buoyantly neutral or unstable to convection, since the resulting change in the vertical disk structure might allow for VSI, even if cooling becomes less efficient.} \\ \indent
\ADD{It has to be noted, that \autoref{VSI_growth} does not include the effects of viscosity which can not be treated extensively in this work. \cite{Lin_2015} found that significant growth in set-ups including viscosity only occurs for modes with $k_R H \sim \mathcal{O}(10)$. We are therefore limiting our study to a radial wavenumber in this order of magnitude to minimise potential shortcomings of the linear theory due to viscous damping in our disk models.}

\subsection{Vertical Convective Instability}\label{subsec:vci}
Vertical convection is not a primary instability mechanism we study for this paper. 
But our 1+1D models do contain internal heating from viscosity which in combination with the radiation 
transport can generate vertical temperature stratification that becomes super-adiabatic for sufficient optical depth.

Vertical buoyancy (Vertical Convective Instability - VCI) is the consequence of such a temperature profile. \cite{Cameron_1978} first considered this to be a possible source of turbulent viscosity in protoplanetary accretion disks. Further study revealed that convection does not significantly influence the angular momentum transport but might be able to establish a vertically adiabatic stratification \citep{Cabot_1987, Klahr_2007}. This makes vertical buoyancy an important aspect of the disk stability, even if it might not be able to drive turbulence in the disk, as it allows for the growth of VSI even if relaxation times exceed the value for the vertically isothermal stratification (\autoref{VSI_crit}). \\ \indent
\ADD{More recent studies, like \cite{Lesur_Ogilvie_2010} have revisited the possibility of vertical convection as a source of angular momentum transport in disks. They indeed found positive $\alpha$ stresses but pointed out that a continuous level of turbulence requires a steady maintenance of the unstable vertical temperature gradient, which in their work was superimposed by the input disk structure. \cite{Held_2018} point out that a secondary mechanism like spiral density waves due to an orbiting planet \citep{ Lyra_2016, Boley_2016} or dissipation of strong magnetic fields, created by Hall-MHD \citep{Lesur_2014} might be able to render the vertical entropy gradients unstable. For such a sustained unstable stratification, they also find substructures like vortices and zonal flows emerging from convective instability. Whether the supporting mechanisms are able to provide the necessary thermal energy at the right places inside of the disk is nonetheless uncertain. The main problem of vertical convection as a source of angular momentum transport thus remains to be the missing ability of the convective instability to self-consistently render the temperature gradient steeper than adiabatic.\\ \indent
The corresponding condition on the vertical stratification is given by the Schwarzschild-Criterion and can also be phrased in terms of a negative vertical entropy gradient \citep[e.g.][]{Pringle_King_2007} }
\begin{equation}\label{conv}
\frac{\partial}{\partial z}C_{\mathrm{V}} \log\left(\frac{P}{\rho^{\gamma}}\right)=\frac{\partial S}{\partial z}<0 \quad 
\Rightarrow \quad \text{Instability},
\end{equation}
where $C_{\mathrm{V}}$ represents the heat capacity at constant volume, $\gamma$ is the heat capacity ratio, and $S$ gives the specific entropy.
The growth rate of the VCI can be determined via the vertical Brunt-V\"ais\"al\"a-frequency $N_z$ \citep{Ruediger_2002}
\begin{equation}\label{vci_growth}
\Gamma_{\mathrm{VCI}}=\sqrt{-N_z^2}=\sqrt{\frac{1}{\gamma \rho} \frac{\partial P}{\partial z} \frac{\partial}{\partial z} \log\left(\frac{P}{\rho^{\gamma}}\right)}\,.
\end{equation}
The violation of the criterion may lead to vertical convective motion, which could be treated in a mixing length model \citep{Bell_1997}, or simply used to limit the 
the vertical temperature gradient to be adiabatic. We do not apply any of these assumptions here for our disk modelling in the following section, because in the interplay with the other instabilities it is unclear how vertical entropy transport will actually be established as long as the relevant numerical simulations have not been performed.

\subsection{Thermal Relaxation}\label{subsec:therm}
The thermal relaxation times of the disks' material determine how fast a temperature perturbation decays or, in other words, how fast a spatially perturbed fluid parcel adopts the temperature of its new surrounding. This makes $\tau_{\mathrm{relax}}$ an important parameter for the COS (see \autoref{COS_rate}), the SBI (see \autoref{SBI_growth}), and the VSI (see \autoref{VSI_crit}), because the growth of these instabilities relies on temperature differences between the perturbed flow and the background.
\citet{Malygin_2017} derived a detailed formalism for the calculation of the thermal relaxation times for the optically thin as well as for the optically thick regime in protoplanetary disks. Our model provides the necessary information to compute these time scales and therefore allows to make statements about the linear growth phase of the instabilities considered here.
The radiative transfer in the optically thick regime is limited by the diffusion of photons which happens on a time-scale
\begin{equation}\label{t_diff}
\tau_{\mathrm{diff}}=\frac{1}{\tilde{D}_{\mathrm{E}} k^2},
\end{equation}
where $k^2 = k_R^2 + k_z^2$ is the wavenumber of the perturbation and 
\begin{equation}\label{D_diff}
\tilde{D}_{\mathrm{E}} = D_{\mathrm{E}}\cdot f =\frac{\xi c}{\kappa_{\mathrm{op}} \rho}\cdot \frac{4\eta}{1+3 \eta}
\end{equation}
represents the effective energy diffusion coefficient, where $\kappa_{\mathrm{op}}$ refers to the opacity, $\xi$ represents the flux limiter to treat the transition from optical thick to thin properly \citep{Levermore_1981}, $c$ is the speed of light and $\eta=E_{\mathrm{R}}/(E_{\mathrm{R}}+E_{\mathrm{int}})$ 
is the ratio between radiation energy density $E_{\mathrm{R}}=a T^4$ to combined radiation and internal energy density $E_{\mathrm{int}}=\rho C_{\mathrm{V}} T$, with the radiation constant $a$ \citep[see][Appendix A, for more detail]{Malygin_2017}.

For the optically thin case most photons are able to directly leave the material, i.e. optical thin cooling.
In that case not only the radiation emitted by the dust grains but also the coupling between dust and gas will determine the loss rate of thermal energy.

The majority of thermal energy is stored in hydrogen molecules and helium atoms, which are the most abundant species in the disk. Those particles have no electric dipole moment, which makes them extremely inefficient coolants. They therefore have to transfer their energy to the emitting species, which are most importantly dust and ice particles. Their inner structure allows for the direct emission of IR radiation via crystal lattice vibrations. 
The second time scale is therefore set by the collisional time scale for gas molecules to collide with dusty and icy grains 
\begin{equation}\label{t_coll}
	\tau_{\mathrm{coll}}=\frac{1}{n \sigma_{\mathrm{c}} v_{\mathrm{coll}}},
\end{equation}
where $n$ refers to the number density of the grains,
with the collisional cross section $\sigma_{\mathrm{c}}$. For this work we adopted the estimates from \cite{Malygin_2017}, who estimated $\sigma_{\mathrm{c}}\approx \SI{1.5e-9}{\square\centi \metre}$. The thermal energy of gas particles, defined via kinetic gas theory, determines the most likely velocity for a molecule (or atom) at gas temperature $T$ 
\begin{equation}
v_{\mathrm{coll}}=v_{\mathrm{g}}=\sqrt{\frac{3 k_{\mathrm{B}} T}{\mu m_{\mathrm{p}}}},
\end{equation}
with the Boltzmann constant $k_{\mathrm{B}}$ and mean molecular mass $\mu m_{\mathrm{p}}$.
The target number density $n$ is calculated for a constant dust to gas ratio of $\epsilon_{\mathrm{d/g}}=0.02$. \ADD{The considered dust grains are $\si{\micro \metre}$ sized, implying efficient coupling between dust and gas, which justifies the assumption of a constant $\epsilon_{\mathrm{d/g}}$ due to efficient mixing.}

If the dust receives the kinetic energy of the gas via these collisions, its emission time-scale is mainly defined by the black-body-emission rate
\begin{equation}\label{t_thin}
	\tau_{\mathrm{emit}}=\frac{C_{\mathrm{V}}}{16 \kappa_{\mathrm{op}} \sigma T^3},
\end{equation}
where $\sigma$ gives the Stefan-Boltzmann constant.
Finally, the total thermal relaxation time scale of the disk's material is determined by the slowest channel of energy transfer
\begin{equation}\label{t_relax}
	\tau_{\mathrm{relax}}=\max{(\tau_{\mathrm{coll}},\tau_{\mathrm{diff}},\tau_{\mathrm{emit}})}.
\end{equation}
This means that the thermal relaxation in the dense inner regions is dominated by the diffusion time scale and the regions farther away from the central star cool mostly via direct irradiation. Since temperatures decline with distance, optically thin cooling becomes less efficient as well, because the black body emission rate scales with $T^{-3}$.
Far above the midplane, where densities nearly reach those of typical molecular clouds, collisions become so unlikely that the energy transfer between hydrogen molecules and the emitting species becomes limited by the collisional timescale, which is completely in line with the arguments in \citet{2018arXiv181006588B}, where this effect is even enhanced by dust growth and sedimentation.

\section{Model}\label{sec:models}
\subsection{Structure Model}
In order to calculate the physical conditions in the $R-z$ plane of a protoplanetary disk, a 1+1D steady state accretion disk model was used \citep{Meyer_1982, DAlessio_1998}. Comparable setups were also implemented by \cite{Bell_1993} and \cite{Bell_1997} in their studies of the structure of equilibrium disks and the evolution and origins of FU-Orionis events. The model used, consists of a radial series of vertical integrations, executed via finite differences in a cylindrical grid. Input parameters are the disk mass $M_{\mathrm{disk}}$, the $\alpha$-parameter and the stellar mass $M_*$. 

Classically these models used the accretion rate $\dot{M}$ as input parameter, because this is the observable that was to be modelled. But in planet formation we are more interested in the disk mass, thus in a work-around we define the desired disk mass and iteratively search for an accretion rate that is consistent with this disk mass for the given $\alpha$ and $M_*$, similar to the method undertaken in \cite{Andrews2010} with the difference that they applied a 1D radial model, whereas we reconstruct the vertical structure as well. To determine the mass of our disk model we use an inner cutoff at $\SI{0.1}{\AU}$ and create an exponential truncation radius of $R_c = \SI{100}{\AU}$ for all disk models, for the sake of a finite disk mass, by
setting $\dot{M}(R) = e^{-R/\SI{100}{\AU}} \dot{M}$ for 
the integration of the structure models,
which leads to a truncation in $\Sigma$ as a result. \ADD{This means the accretion rate drops by $\sim 50\%$ at $\SI{60}{\AU}$ thus mimicking the results of \cite{LyndonBellPringle_1974} for viscous ring spreading, implying a radial decline of accretion rates towards the truncation radius and negative accretion rates at larger radii and thus outside of the regions we are investigating. The outward movement of material at large radii is a necessary consequence of angular momentum conservation in accretion disks and does not imply an unstable disk (whereas an accretion disk is always just quasi steady by nature, which means that the viscous time scale is much larger than the dynamical timescale).}
Here we ignore for the time being that the disk truncation radius might also vary as function of stellar and disk masses.
Thus, we define
\begin{equation}
M_{\mathrm{disk}} = 2 \pi \int_{\SI{0.1}{\AU}}^{\SI{100}{\AU}} \Sigma{(R)} R \, \mathrm{d}R,
\end{equation}
using the $\Sigma(R)$ values as determined in our model.

When on the main sequence, then stellar mass also defines the star's radius ($R_*$), effective temperature ($T_{\mathrm{eff}}$) and luminosity ($L_*$) via standard mass-radius and mass-luminosity relations \citep{Duric_2003, Salaris_2005, Weigert_2009} (see \autoref{tab:stellar}). 
For pre-main sequence stars that are the ones around which we find disks the luminosity can be up to $50\%$ larger \citep{1998A&A...337..403B}, an effect that we neglect for the present paper as it will only slightly increase the temperature of the disk, thus further increase the likely-hood for thermal driven instabilities. 

For each vertical disk column we solve 
locally for the condition of vertical hydrostatic equilibrium 
\begin{equation}\label{pgrad}
\frac{\mathrm{d} P}{\mathrm{d} z}=\rho g_z=\frac{\rho G M_* z}{(R^2+z^2)^{\nicefrac{3}{2}}},
\end{equation}
where $G$ gives the gravitational constant. 
The dissipation of kinetic energy, caused by the $\alpha$-viscosity \citep{Shakura_1973} in a slight modified way to include the vertical stratification of the disk is given by
\begin{equation}
\nu = \alpha c_s \min{\left(H, l_P\right)},
\end{equation}
where $H = c_s/ \Omega$ is the standard pressure scale height in the midplane and $l_P = P/(\partial_z P)$ is the local pressure scale height, to limit dissipation far above the midplane.
Viscous heating is then equated with the gradient of radiative flux
\begin{equation}\label{fgrad}
\frac{\mathrm{d} F}{\mathrm{d} z}=Q^+=\frac{9}{4} \rho \nu \Omega^2,
\end{equation}
in order to obtain vertical thermal balance.
The resulting energy flux is furthermore associated with a temperature gradient via the flux limited diffusion equation by \citet{Levermore_1981}, which is used to obtain the disk's temperature profile
\begin{equation}\label{tgrad}
\frac{\mathrm{d}T}{\mathrm{d} z}=\frac{F}{D_{\mathrm{T}}}=\frac{\rho \kappa_{\mathrm{op}}}{4 c \xi a T^3}F.
\end{equation}
The gas density is calculated with the ideal equation of state $P=R \rho T/\mu$, with gas constant $R$ and mean molecular weight $\mu = 2.33$.
Equations \eqref{pgrad}, \eqref{fgrad} and \eqref{tgrad} are integrated vertically top down towards the midplane by use of finite differences \citep{Bell_1997}. 
This kind of model does not allow for the modelling of radial energy transport, which is an appropriate approximation as long as the disk remains geometrically thin \citep{Pringle_1981}.
In contrast to the model by \cite{Bell_1993}, our  calculations start at the highest grid-cell of the simulation domain, arbitrarily chosen, and not at a cell which full-fills $\tau=\nicefrac{2}{3}$. Therefore the initial temperature is assumed to be equal to the temperature caused by stellar irradiation alone
\begin{equation}
T_{\mathrm{init}}=T_{\mathrm{eff}} \sqrt{\frac{R_*}{R}}\sin{(\theta)}^{\nicefrac{1}{4}}
\end{equation}
where $\theta$ corresponds to the approximate angle between the disk's surface and the line of sight. 
In \cite{Bell_1997} we still had to do two independent integration to cover the optical thick and thin parts of the disk. One going down from the $\tau=\nicefrac{2}{3}$ photosphere and one going up. This is now combined in one integration sweep.

The initial vertical energy flux is defined via the equilibrium condition for actively accreting disks \citep{Pringle_1981}
\begin{equation}
F_{\mathrm{init}}=\frac{3GM_* \dot{M}}{8 \pi R^3}\left(1-\sqrt{\frac{R_*}{R}}\right).
\end{equation}
This equation links the initial guess of the disk's mass accretion rate $\dot{M}$ with the surface energy flux of a disk annulus.
A vertical integration series is finished when the initial density guess $\rho_{\mathrm{init}}$ leads to a vanishing flux in the disk midplane. Otherwise, $\rho_{\mathrm{init}}$ is varied and the vertical integration is rerun.
This procedure is repeated at every radial position, in order to obtain the whole $R-z$ plane structure.
The radial series is complete
when the resulting disk mass fulfils $|M_{\mathrm{disk}}-M_{\mathrm{d,res}}|/M_{\mathrm{disk}}<10^{-4}$, where $M_{\mathrm{disk}}$ is the input parameter of the model and $M_{\mathrm{d,res}}$ is the resulting disk mass of a radial integration series.

If this condition is not fulfilled, $\dot{M}$ is varied and the whole radial process is repeated until the disk mass fits the demanded value.
The resulting disk structure is then used to calculate the analytical growth rates and instability criteria introduced in \autoref{sec:instabilities} for the whole $R-z$ plane.

\subsection{Opacity Model}
\begin{figure}[ht!]
\centering
\includegraphics[width=0.455\textwidth]{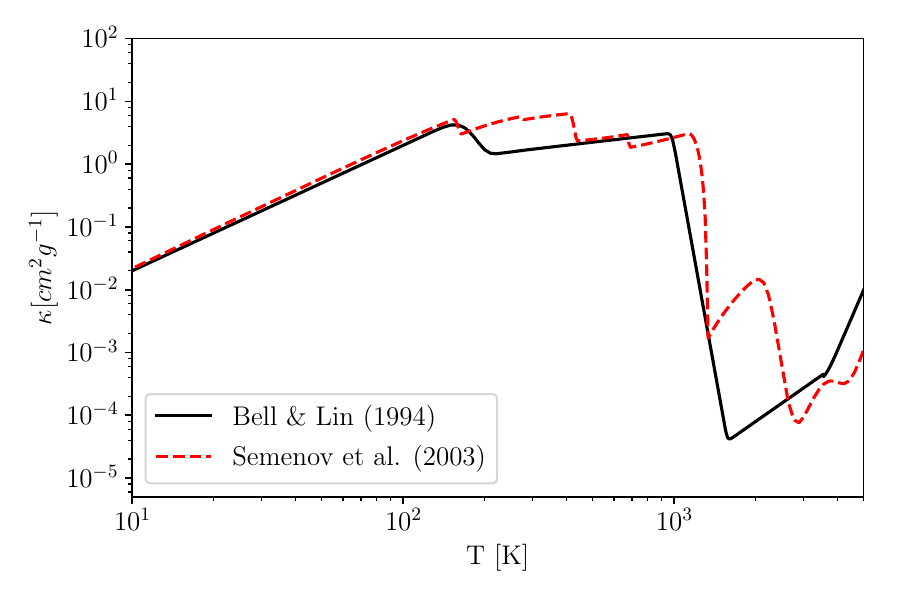}
\caption{\citet{Bell_1993} Rosseland mean opacities, used in the scope of this work, for different temperatures and a given density of \SI{e-9}{\gram \per \cubic \centi \metre } compared to the more recent, but very similar opacity model by \cite{Semenov_2003}.}
\label{opa}
\end{figure}

The assumption of vertical thermal balance requires radiative transfer in the $z$-direction, which is realised via the flux-limited diffusion approach by \citet{Levermore_1981}. The underlying temperature diffusion coefficient depends on the opacity structure $\kappa_{\mathrm{op}}(R,z)$ via
\begin{equation}
D_{\mathrm{T}}=\frac{4 c \xi a T^3}{\rho \kappa_{\mathrm{op}}}.
\end{equation}
The applied opacity model is therefore of great importance for the disk structure and its hydrodynamic stability, because it determines the magnitude of temperature gradients. 
For the calculations done in the scope of this work, the opacity model by \cite{Bell_1993} was used, which relies on the analytical expressions by \citet{Lin_1980}, who assumed grain sizes in the \si{\micro \meter} range. It provides frequency independent mean opacities for eight different regions of protoplanetary disks, defined and ordered by their temperature.
In each region, opacities are calculated by a specific power law in temperature and density
\begin{equation}
\kappa_{\mathrm{op},i}=\kappa_{\mathrm{op}_ 0,i}\rho^{a_{\mathrm{i}}} T^{b_{\mathrm{i}}}.
\end{equation} 
Region 1 contains ice grains and metal grains, the opacity in region 2 is determined by the evaporation of the ice grains and the still existing metal grains. The regions 3 and 4 are defined by the abundance of metal grains and their evaporation. At higher temperatures, molecular hydrogen dominates region 5 until hydrogen scattering determines opacities in region 6. When the gas is ionised, electron scattering and Kramer's law take over in regions 7 and 8.
In our model, temperature do not exceed $\sim \SI{1800}{\kelvin}$, which means that only the opacities of region 1-5 matter for our considerations.
Metal and water ice grains therefore determine the opacity in our model, which leads to two major drops in \autoref{opa} at the typical evaporation temperatures (water ice at $\sim \SI{160}{\kelvin}$; metal/silicate grains at $\sim \SI{1000}{\kelvin}$). We will see that the transition zones of opacity are the prime locations for radial buoyancy driven instabilities, because the temperature dependent opacities 
generate major fluctuations in the radial midplane temperature profile (see \ref{Midplane}), while leaving the average profile unchanged.

\subsection{Radial Transport of Heat}
\begin{figure*}[ht]
\includegraphics[width=1.0\textwidth]{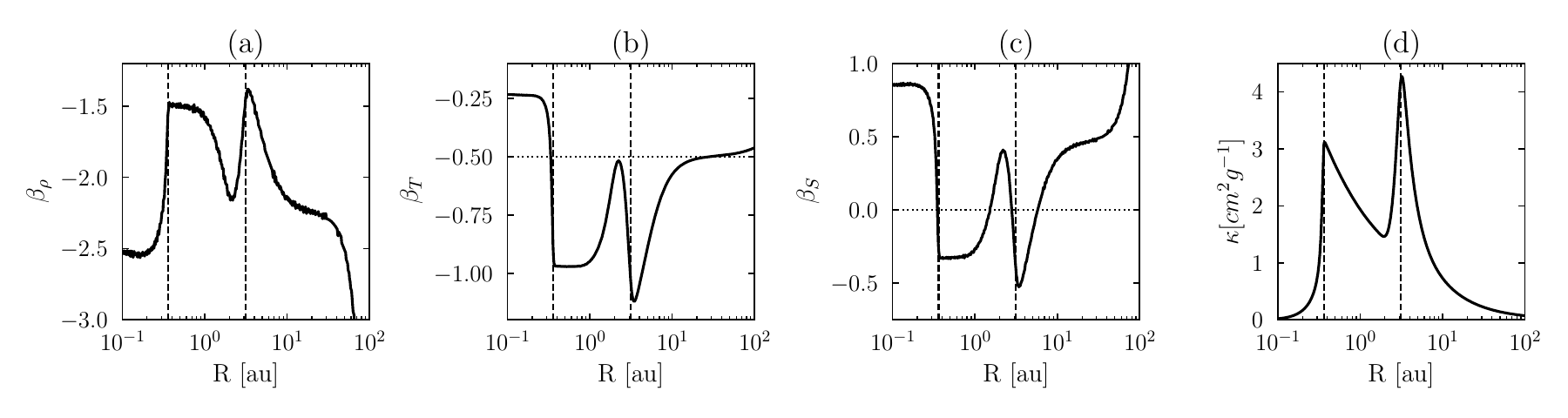}
\caption{Logarithmic midplane gradients of density (a, $\beta_{\rho}=\frac{\mathrm{d} \log (\rho)}{\mathrm{d} \log (R)}$), temperature (b, $\beta_{T}=\frac{\mathrm{d} \log (T)}{\mathrm{d} \log (R)}$) and entropy (c, $\beta_S=\beta_T+(1-\gamma)\beta_{\rho}$) \citep{Klahr_2004}, as well as the midplane opacity profile (d) for a solar mass star with a disk of \SI{0.1}{\solmass} and $\alpha=10^{-3}$. The vertical dashed lines represent locations of the two grain evaporation lines (water ice at $\sim \SI{160}{\kelvin}$; metal grains at $\sim \SI{1000}{\kelvin}$) in the midplane. It can be seen that the evaporation of ice grains 
and metal grains create regions where the entropy gradient is negative for a certain radial range, whereas the global gradient remains positive on the global average.}
\label{Midplane}
\end{figure*}

Our 1+1D model consists of independent vertical slices of the disk. Thus radial transport of heat via radiation transport is neglected as is radial transport of entropy via local mass advection. Both could have an influence on the radial temperature structure and thus on the entropy profile, which is so important for the onset of the convective instabilities. \cite{Bitsch2015} created a model for an evolving disk around a solar mass star for an $\alpha = 5.4 \times 10^{-3}$, which used a full 3D radiation hydrodynamical simulation in axissymmetry.
Despite their proper treatment of radial diffusion of radiation, they find the same peak values of the temperature gradient in the evaporation zone of the ice particles, which they fit with $\beta_T= -\nicefrac{8}{7}$, which is the same value that we find in our much simpler simulations (See Fig.\ \ref{Midplane}b).

\section{Results}\label{sec:results}
In order to probe the parameter space of star-protoplanetary disk systems, we calculated a series of structure models for different values of each parameter ($M_{\mathrm{disk}}, M_*, \alpha$). See \autoref{tab:params} for an overview of our simulation parameters, including the mass accretion rate for each model.
We do not change the assumed metallicity from the solar value assumed in \cite{Bell_1993}, which would reflect in higher or lower opacities. In that sense choosing a lower disk mass (compensated by slightly larger $\alpha$ to achieve the same accretion rate) would have a similar effect on the disk structure as decreasing the opacity. Nevertheless investigations of the effect of metallicity should eventually be done in the context of better dust opacities, including the evolution of the dust population as in \cite{Birnstiel2012}. \cite{Estrada2016} have put forward a model in which disk and dust are evolving and opacities are calculated from the local dust properties, yet their model does not determine the detailed vertical structure of disks, thus does not derive the local radial and vertical stratification of their disk as function of $R$ and $z$.

\begin{deluxetable}{cccc}[ht]
\tablecaption{Stellar Parameters mass, luminosity, effective temperature and radius used for the models presented in this work.
\label{tab:stellar}}
\tablecolumns{4}
\tablenum{1}
\tablewidth{0pt}
\tablehead{
\colhead{$M_* [\si{\solmass}]$} &
\colhead{$L_*$ [\si{\sollum}]} & \colhead{$T_{\mathrm{eff}} \mathrm{[T_{\odot}]}$ } & \colhead{$R_* \mathrm{[R_{\odot}]}$}
}
\startdata
0.4	 & 	0.028 & 0.647 & 0.4 \\
0.6 & 0.13 & 0.775 & 0.6 \\
1.0 & 1.0 & 1.0 & 1.0 \\
1.5 & 5.063 & 1.328 & 1.275 \\
\enddata
\end{deluxetable}

\begin{deluxetable}{cccr}[ht]
\tablecaption{Parameters and mass accretion rates of the structure models presented in this work. Stellar mass, disk mass and $\alpha$ are the input parameters of the model. The mass accretion rate is determined iteratively in order to fit the input parameters.
\label{tab:params}}
\tablecolumns{4}
\tablenum{2}
\tablewidth{0pt}
\tablehead{
\colhead{$M_* [\si{\solmass}]$} &
\colhead{$M_{\mathrm{disk}} [\si{\solmass}]$} & \colhead{$\alpha$} & \colhead{$\dot{M} [\si{\solmass \years^{-1}}]$}
}
\startdata
0.4	 & 	0.1	 & 	0.001	 & 	$7.7282\times 10^{-9}$	 \\ 
0.6	 & 	0.1	 & 	0.001	 & 	$8.4444\times 10^{-9}$	 \\ 
1.0	 & 	0.1	 & 	0.001	 & 	$1.0513\times 10^{-8}$	 \\ 
1.5	 & 	0.1	 & 	0.001	 & 	$1.27555\times 10^{-8}$	 \\ 
\midrule 
1	 & 	0.01	 & 	0.001	 & 	$1.0304\times 10^{-9}$	 \\ 
1	 & 	0.05	 & 	0.001	 & 	$5.2031\times 10^{-9}$	 \\ 
1	 & 	0.1	 & 	0.001	 & 	$1.0513\times 10^{-8}$	 \\ 
1	 & 	0.2	 & 	0.001	 & 	$2.13935\times 10^{-8}$	 \\ 
\midrule 
1	 & 	0.1	 & 	0.00001	 & 	$1.03\times 10^{-10}$	 \\ 
1	 & 	0.1	 & 	0.0001	 & 	$1.0361\times 10^{-9}$	 \\ 
1	 & 	0.1	 & 	0.001	 & 	$1.0513\times 10^{-8}$	 \\ 
1	 & 	0.1	 & 	0.01	 & 	$1.088503\times 10^{-7}$ \\
\enddata
\end{deluxetable}

\begin{figure*}[ht] \centering
\includegraphics[width=1.0\textwidth]{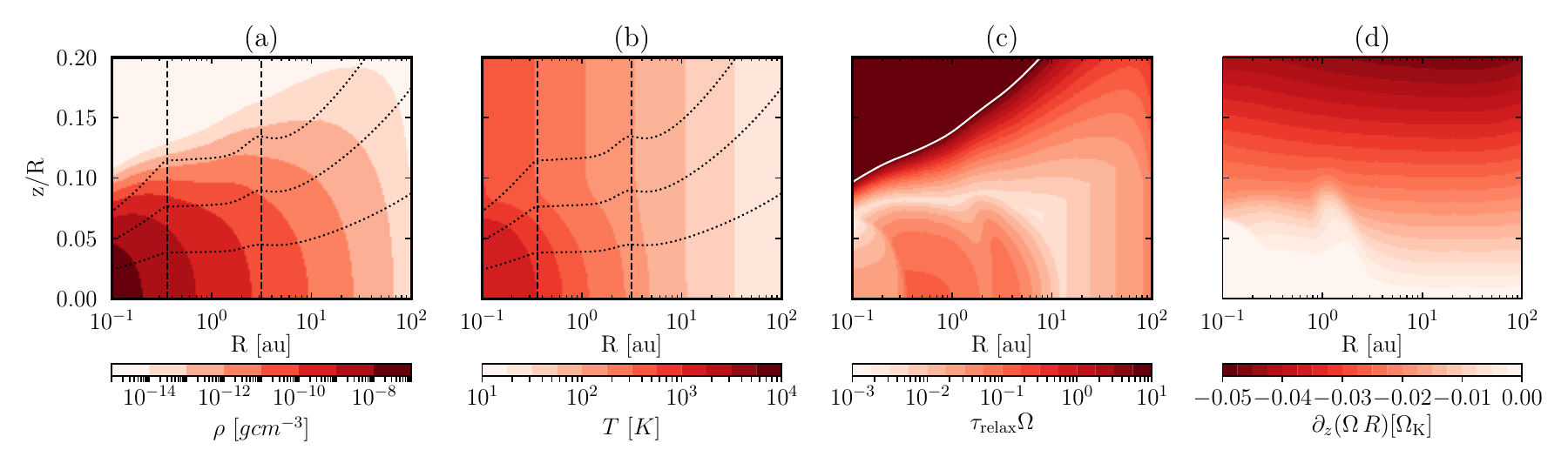}
\caption{Densities (a), temperatures (b), thermal relaxation times for a wavenumber of $k=50/H$ (c) \ADD{and vertical shear (d)} , calculated with the 1+1D steady state model for the input parameters $M_{\mathrm{disk}}=0.\SI{1}{\solmass}$, $M_*=\SI{1}{\solmass}$, $\alpha=10^{-3}$. Dotted lines indicate 1, 2 and 3 pressure scale heights. The kinks occurring in the scale height profiles arise because of the maxima in opacity, shown in  \autoref{Midplane}d. The white line in (c) indicates $\tau_{\mathrm{relax}} \Omega =10$, which is only reached high above the midplane, because low densities reduce the frequency of collisions and therefore hamper the energy transfer onto to emitting species.}
\label{Profiles}
\end{figure*}

\subsection{Disk Structure and Stability}\label{subsec:structure}
Far away from the central object ($R\gtrsim \SI{10}{\AU}$), densities are low and 
stellar irradiation dominates the thermal structure, which leads to a radial trend in temperature that scales with $R^{-0.5}$ (horizontal line in \autoref{Midplane}b). 
The temperature profile closer to the star, which is strongly influenced by the disk's varying optical depth structure has a major impact on thermally driven instabilities. It can be seen in \autoref{Midplane}c, that entropy gradients drop below zero for the most opaque zones, which renders the disk radially buoyant (the pressure gradient is negative here) in the sense of the classical Schwarzschild Criterion ($\partial_R S \partial_R P>0$) \citep[e.g.][]{Pringle_King_2007}, which is the necessary condition for the onset of COS. 
Two dimensional density and temperature structures are shown in \autoref{Profiles}. As mentioned before, 
variations in the opacity (mostly with temperature) lead to complex behaviour of the density and temperature gradients. 
Notably one finds the steepest temperature gradient and the most shallow density gradient at the grain and ice evaporation lines (indicated by vertical lines) at which opacity and thus optical depth reaches a local extremum.
The anticorrelation of temperature and density gradients is typical for viscous accretion disks, because they have a roughly constant pressure profile of about $R^{-\nicefrac{3}{2}}$, which is a result from the $\alpha-$model for viscosity and the assumption of a constant accretion rate.
Temperature is then also relatively high at these locations, because thermal energy transport needs stronger gradients with increasing optical depth. Therefore, the disk appears to be puffed up in the regions of maximal opacity which can be seen as the kinks in the scale height profiles in \autoref{Profiles}. The two dimensional temperature profile also shows strong vertical gradients in the viscously heated parts of the disk and a vertically isothermal structure in the regions dominated by stellar irradiation. The vertical gradients are of interest for the investigation of 
VSI. 

The outer, vertically isothermal regions are prone to be unstable due to the VSI, because 
cooling times are low enough to allow for the growth of vertical perturbations. \autoref{Profiles}c displays the thermal relaxation times for a fixed perturbation wavenumber of $k=50/H$ (which was used as the radial wavenumber for the study of the VSI). It can be seen that relaxation times are low at distances $\gtrsim \SI{10}{\AU}$, which is a necessary condition for VSI growth. Additionally, the criterion $\tau_{\mathrm{relax}} \Omega<10$ \citep{Nelson_2012} for a polytropic stratification is fulfilled  in those regions that are prone to be convective. This means, that the VSI can also exist in the dense interior regions if the disk becomes buoyant there and a polytropic structure is established. 

\ADD{The disk's vertical gradient in angular velocity is shown in \autoref{Profiles}d, where it can be seen that the vertical shear increases with height above the midplane. It's origin is the vertical variation of the radial pressure gradient. Radial hydrostatic equilibrium, i.e. the balance of gravitational force vs. radial pressure gradient and centrifugal force, thus requires the rotation velocity to change too. This vertical shear is the necessary condition for the VSI.}

\ADD{The disks' stability in the context of self gravity was checked by use of the Toomre-Criterion $Q=\kappa_R C_s/(\pi G \Sigma)<1$, \citep{Toomre_1964} where $C_s=\partial \tilde{P}/\partial \Sigma$ represents a two dimensional speed of sound \citep{Pringle_King_2007}. Gravitational stability holds for the disk models, presented in this work (for more detail, see Appendix, \autoref{Toomre}). We find models only becoming gravitationally unstable at radii larger than $\sim$\SI{50}{\AU} for a setup of $M_*=\SI{1}{\solmass}$, $M_{\mathrm{disk}}=\SI{0.5}{\solmass}$ and $\alpha = 10^{-3}$, which are therefore not presented here.}

\subsection{Convective Overstability}\label{subsec:cos_res}
\begin{figure}[ht] \centering
\gridline{
\fig{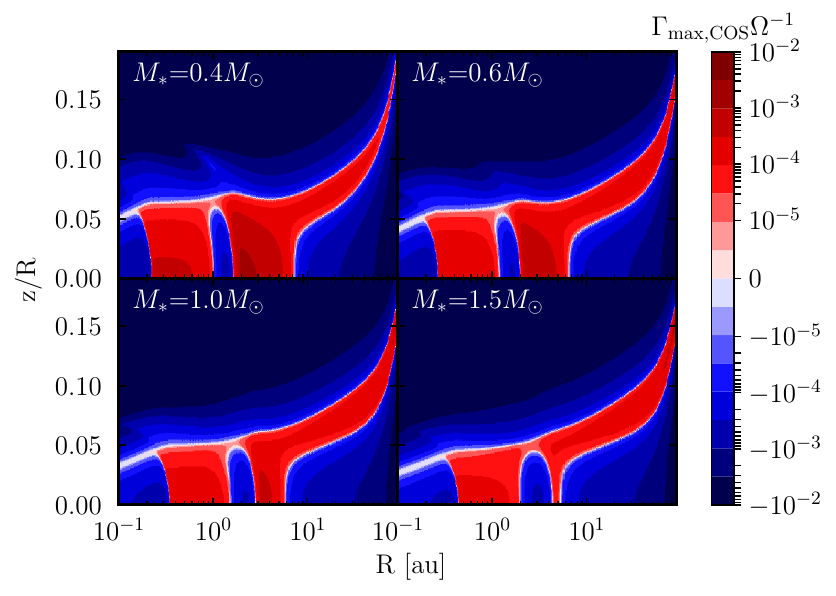}{0.455\textwidth}
{(a) maximal COS growth rates for different stellar masses}}
\gridline{
\fig{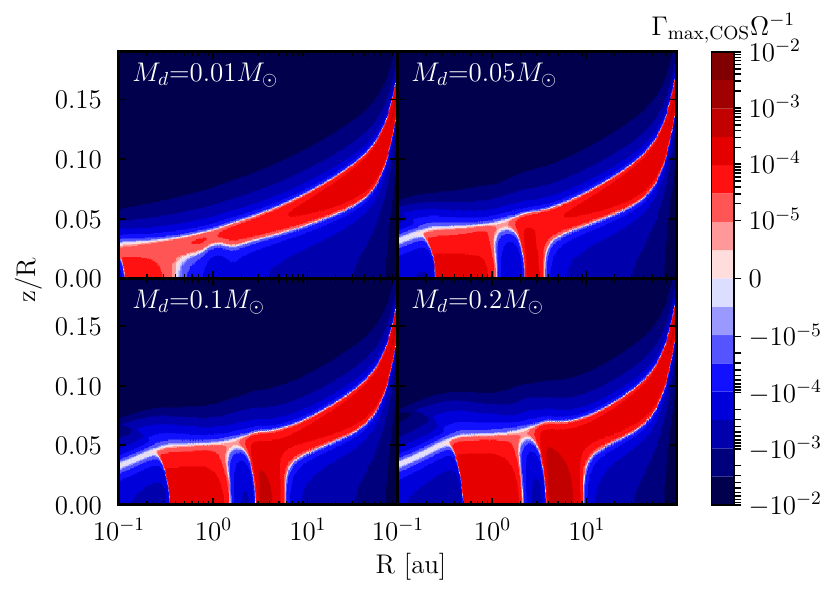}{0.455\textwidth}
{(b) maximal COS growth rates for different disk masses}}
\gridline{
\fig{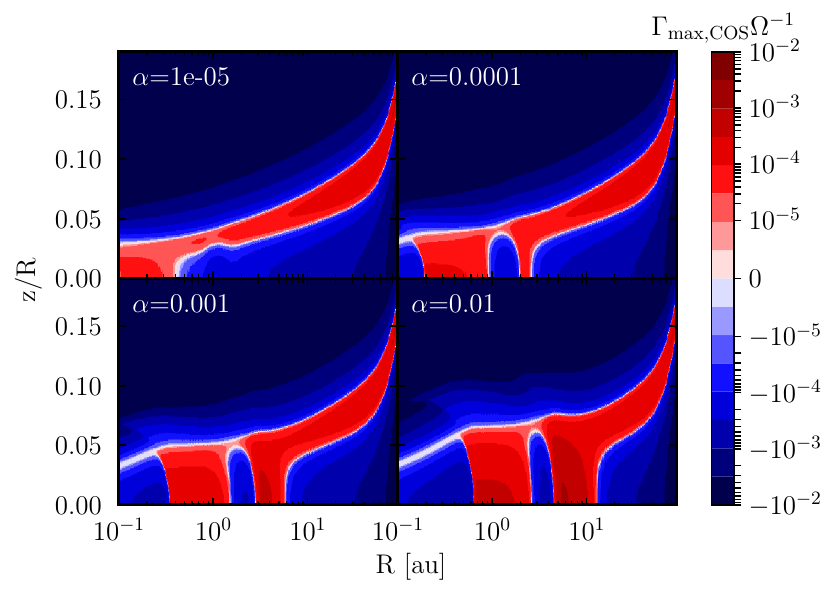}{0.455\textwidth}
{(c) maximal COS growth rates for different $\alpha$-parameters}}
\caption{Maximal growth rates of the linear COS in units of the local dynamical time-scale ($\Gamma \Omega^{-1}$). One parameter is varied per panel, the other two are held constant at $M_*=\SI{1}{\solmass}$, $\alpha=10^{-3}$ or $M_{\mathrm{disk}}=\SI{0.1}{\solmass}$.}
\label{COS}
\end{figure}

The COS requires negative radial entropy gradients. 
Sufficient optical depth is needed for a disk to develop super-adiabatic radial temperature gradients. This means that two COS-active regions are located in the regions of maximal opacity. An additional unstable zone arises at a certain height above the midplane ($z\sim \SI{1}{\scaleheight}$) and covers the whole radial extent of the disk. The reason for this is the change of sign in the radial entropy and pressure gradients in the disk, since there is always a region where the gradients are parallel. This unstable branch therefore also exists in disks without internal heating, as long as radial gradients in temperature exist.
\CORR{As was pointed out in \autoref{subsec:cos}, growth rates reach their maximum for relaxation times that fulfil \autoref{COS_max}, thus defining a maximum growing wavenumber via combination of \autoref{COS_max} and \autoref{t_diff}. For these wavenumber, we find growth rates of $\sim$ \SIrange{e-4}{e-3}{\om}. The dependency of the growth rates magnitude on cooling times is therefore given by the wavenumber, which means that perturbations associated with large $k$ grow better in regions of larger optical depth where cooling needs to be more efficient and small $k$ perturbations grow fastest in regions of smaller optical depth where too fast cooling would render perturbations almost isothermal. This means that the maximum growing perturbations need to be spatially small (large $k$) close to the midplane and become larger (smaller $k$) in the optically thinner upper and outer parts of the disk (see Appendix \autoref{COS_k}). \\ \indent
Viscosity and its impact on small perturbations, as discussed in \cite{Klahr_Hubbard_2014} and \cite{Latter_2016} was not taken into account for our study. \\ \indent
}
As can be seen in \autoref{COS}a, a higher stellar mass (and luminosity) leads to a thermal structure in which the influence of accretion heating becomes less dominant. Therefore, entropy gradients are negative in large regions for low mass stars with massive disks and smaller for high mass stars with a disk of low mass compared to the stellar mass. 
Another important effect of an increased solar luminosity is the overall temperature rise of the disk. Ice sublimates at $\sim$ \SI{160}{\kelvin} and therefore the second COS active zone vanishes for high mass stars (lower right panel in \autoref{COS}a), since temperatures are too high to allow for the existence of opacity enlarging ice grains.

The opposite effect is visible for a larger disk mass (\autoref{COS}b). Because of the increase of accretion heating and optical depth with this parameter, temperature and entropy gradients are both stronger. This also leads to the outwards shift of the susceptible regions with increasing disk mass, which is the direct consequence of the movement of the dust and ice sublimation lines. For a disk of \SI{0.01}{\solmass}, the ice sublimation line is so close to star, that no second unstable zone exist (upper left panel in \autoref{COS}b). \CORR{Due to the increasing densities, wavenumbers need to be small in order to allow for more efficient cooling. (see Appendix \autoref{COS_k}b)}

An increase of the disk's $\alpha$-parameter leads to enhanced viscous heating of the dense interior parts. Therefore, also radial temperature gradients are building up and growth rates for the COS become larger when viscous heating dominates the interior temperature profile.

Since the density profile is only slightly altered by the $\alpha$ parameter, thermal relaxation stays relatively constant in the outer regions, where the opacity is provided by ice and metal grains. In the zones closer to the star, temperatures increase due to the increased accretion heating and sublimation lines move outward with increasing $\alpha$.
\ADD{For an extremely low $\alpha=10^{-5}$, we notice that midplane temperatures are low enough to allow for the existence of metal grains even at radii of $\sim \SI{0.1}{\AU}$. Densities in these regions are high thus leading to high opacities and therefore to an optically thick structure. At these locations, perturbation wavenumbers need to be very high in order to allow for efficient enough cooling (see Appendix \autoref{COS_k}c).}

\subsection{Subcritical Baroclinic Instability}\label{subsec:sbi_res}

\begin{figure*}[ht]
\includegraphics[width=1.0\textwidth]{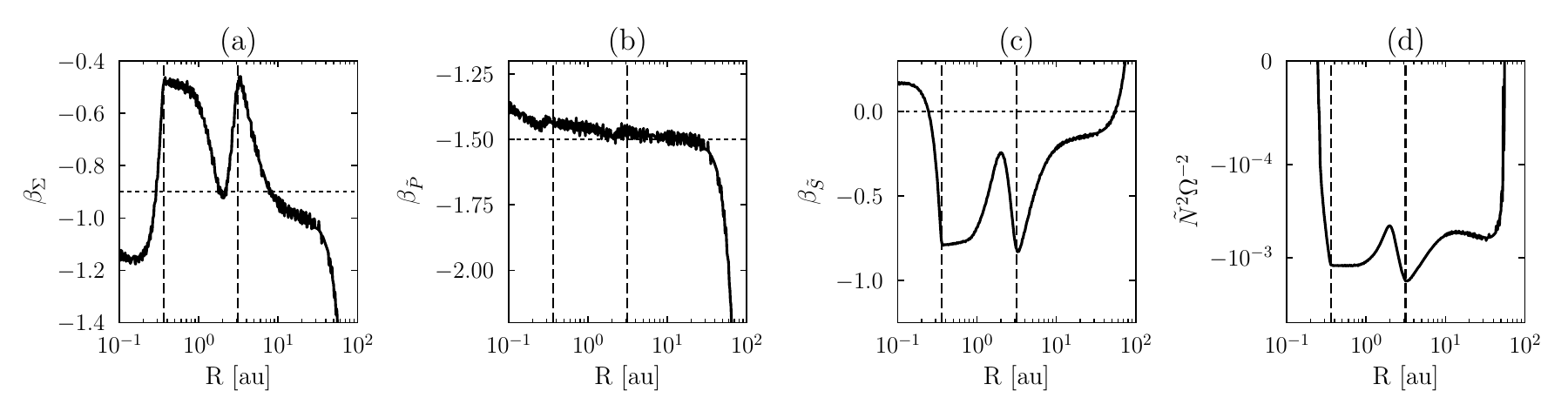}
\caption{Vertically integrated logarithmic midplane gradients of column density (a), pressure (b) and specific entropy (c) as well as the vertically integrated radial buoyancy frequency profile (d) for a solar mass star with a disk mass of \SI{0.1}{\solmass} and $\alpha=10^{-3}$. The vertical lines represent locations of the two grain evaporation lines (water ice at $\sim \SI{160}{\kelvin}$; metal grains at $\sim \SI{1000}{\kelvin}$). The vertically integrated specific entropy gradients expose, that the disk is almost everywhere buoyant for vertically integrated structures like large scale $R-\phi$ vortices.}\label{2DMidplane}
\end{figure*}

\autoref{2DMidplane} displays the quantities, introduced in \autoref{subsec:sbi}. It can be seen that the vertically integrated pressure profile (b) scales with $R^{-\nicefrac{3}{2}}$. The reason for this is the spatially constant mass accretion rate, which means that $\dot{M}\propto \nu \Sigma= \alpha c_s^2 \Sigma \Omega^{-1}  = \alpha \tilde{P} \Omega^{-1}=\mathrm{const.}$ \citep{LyndonBellPringle_1974}. \autoref{2DMidplane}c shows a single large buoyancy unstable region in contrast to the two smaller zones found in the three-dimensional profiles in \autoref{subsec:cos_res}. This means that vertically extended $R-\phi$ vortices, which can be treated as two dimensional structures, might survive in even larger regions than the linear $R-z$ modes of the COS. Migrating vortices could therefore be formed in the COS-active regions and move to the regions unaccessible for small scale COS turbulence.

\begin{figure*}[ht] \centering
\includegraphics[width=1.0\textwidth]{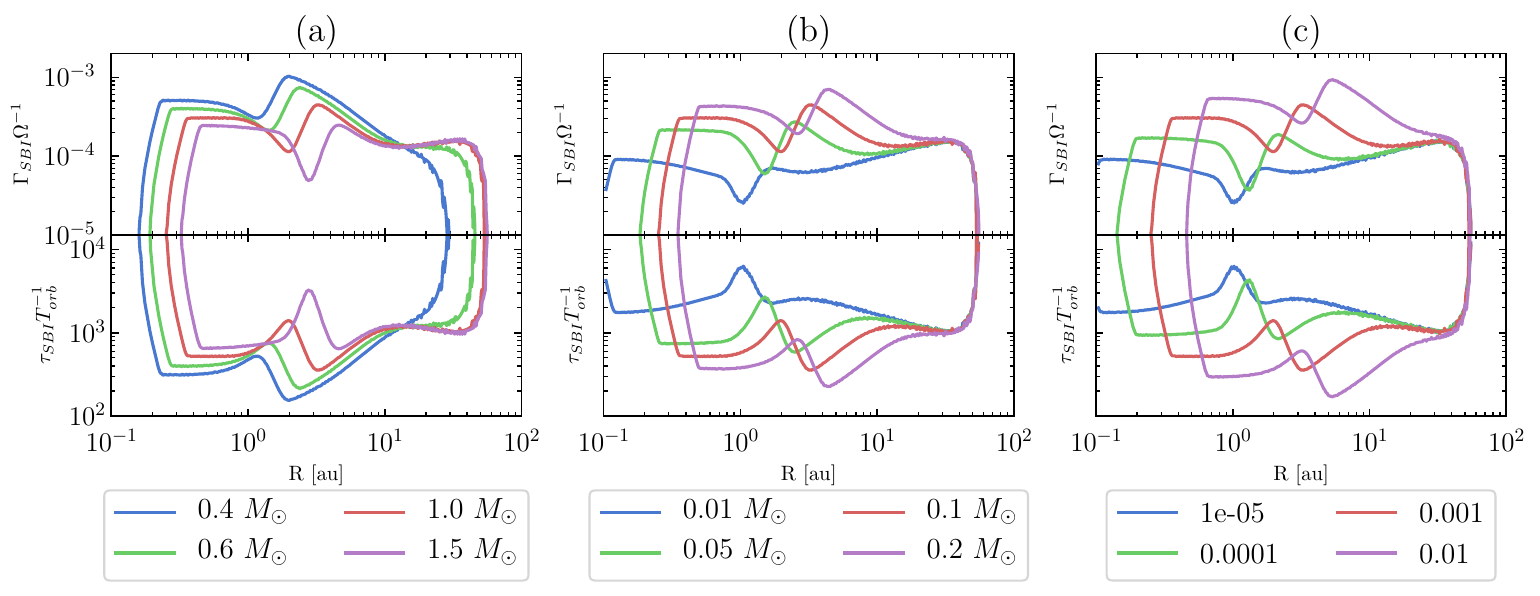}
\caption{Growth rates of an SBI vortex with an aspect ratio of $\chi=4$ for different sets of parameters (top row). The corresponding growth timescales in units of $T_{\mathrm{orb}}=2 \pi/ \Omega$ are show in the bottom row. In each column, one parameter is varied: (a): stellar mass, (b): disk mass, (c): $\alpha$-parameter. The remaining two parameters are set to constant values of $M_*=\SI{1}{\solmass}$, $\alpha=10^{-3}$ or $M_{\mathrm{disk}}=\SI{0.1}{\solmass}$.}
\label{SBI}
\end{figure*}

A parameter study reveals, that the SBI's growth rate has very similar dependency on the disk and stellar parameters as the COS. \autoref{SBI} provides an overview of the SBI's behavior for different parameters. The growth rates shown in \autoref{SBI}a shrink down with increasing stellar mass. The reason for this lies in the decline of radial temperature and entropy gradients in the regions of high opacity as a result of outwards movement of dust evaporation lines, which shifts the optically thick regions into zones of lower density and pressure.
The opposite effect results from an increase of disk mass (\autoref{SBI}b) and $\alpha$-parameter (\autoref{SBI}c), which lead to an increase of accretion heating and hence enlarge radial temperature gradients and vortex growth rates. 
As mentioned before, SBI growth occurs over almost the whole radial extend of the disk, but the highest growth rates are reached at the location of the ice line, where opacities are maximal. At this location, $\Gamma_{\mathrm{SBI}}$ is in the order of $\sim \SIrange{e-4}{e-3}{\om}$, which corresponds to a growth time scale of $\sim \SIrange{e2}{e3}{\orb}$ (local orbital timescales).

\subsection{Vertical Convective Instability}\label{subsec:vci_res}
The onset of VCI in a protoplanetary disk requires the existence of sufficiently steep vertical temperature gradients.
As soon as these gradients are at least adiabatic, the VSI can operate despite long cooling times, which drives our interest in the onset of VCI.

\begin{figure}[ht] \centering
\gridline{
\fig{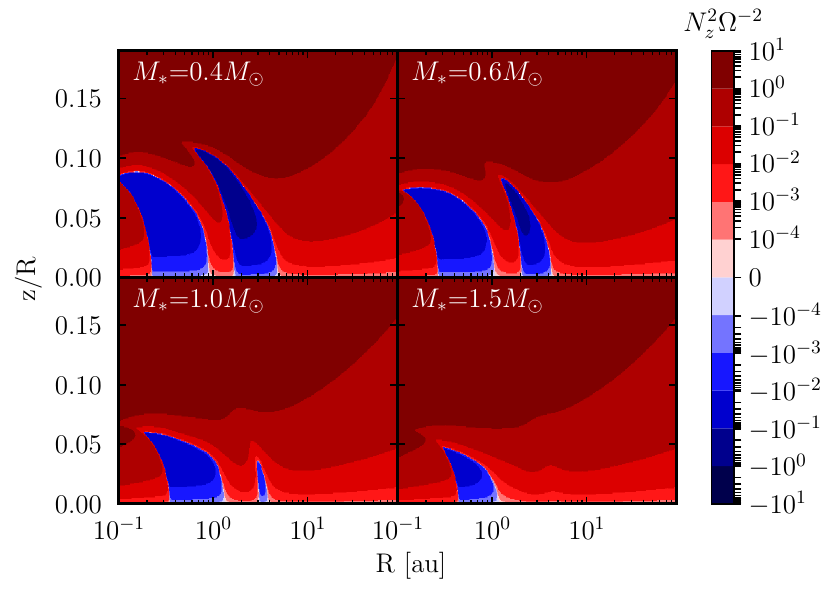}{0.455\textwidth}
{(a) $N_z^2/\Omega^2$ profiles for different stellar masses}}
\gridline{
\fig{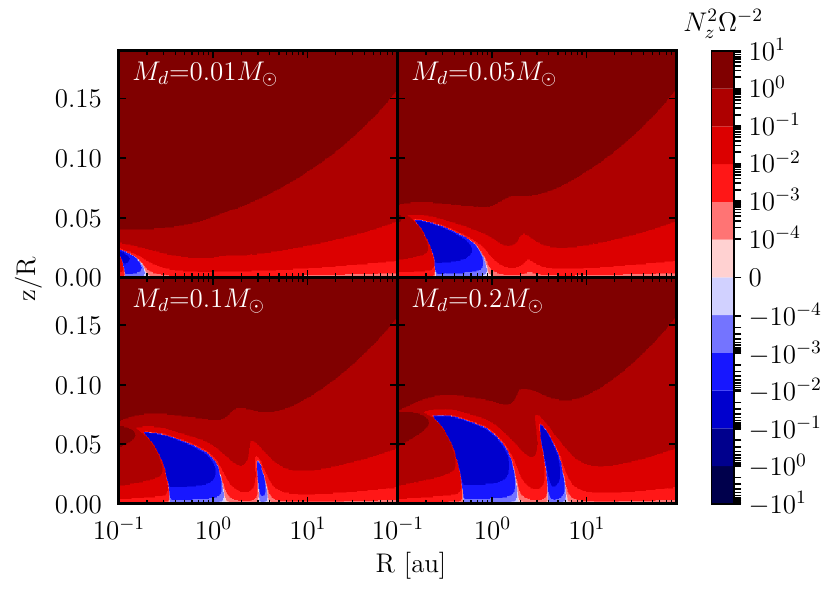}{0.455\textwidth}
{(b) $N_z^2/\Omega^2$ profiles for different disk masses}}
\gridline{
\fig{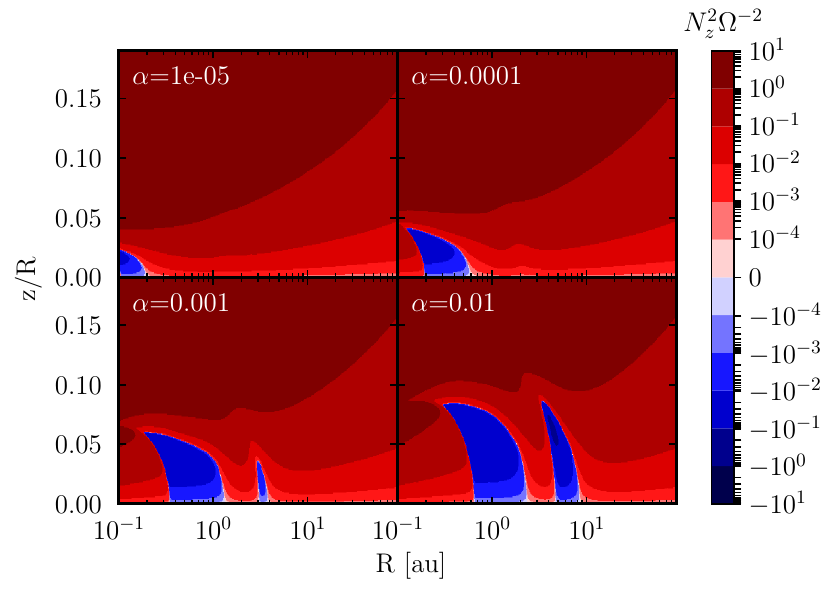}{0.455\textwidth}
{(c) $N_z^2/\Omega^2$ profiles for different $\alpha$-parameters}}
\caption{Squared Vertical Brunt-V\"ais\"al\"a-frequencies in units of $\Omega^2$. Blue color indicates a convectively unstable stratification. One parameter is varied per panel, the other two are held constant at $M_*=\SI{1}{\solmass}$, $\alpha=10^{-3}$ or $M_{\mathrm{disk}}=\SI{0.1}{\solmass}$.}
\label{VCI}
\end{figure}

To steepen the vertical temperature gradients, radiative transfer needs to be inefficient enough to force the temperature to increase close to the midplane. In other words: densities and opacities need to be large and the accretion rate needs to be strong to ensure strong viscous heat production and corresponding temperature gradients.
A temperature profile dominated by stellar irradiation is vertically isothermal and thus contradictory to the conditions for convection. Higher temperatures due to stellar irradiation lead to the evaporation of dust grains at larger radii and to an outwards shift of convection zones.
These effects are strongly visible in \autoref{VCI}a. For these models, disk mass and $\alpha$-parameter are set to constant values of $M_{\mathrm{disk}}=0.\SI{1}{\solmass}$ and $\alpha=10^{-3}$. Two separate convection zones appear for stellar masses $\lesssim \SI{2}{\solmass}$, due to the drop in opacity at $T \sim \SI{160}{\kelvin}$. When the convection zones are shifted to larger radii, densities and rotation frequencies are no longer high enough to sustain the existence of the outer convection zone. An increase of stellar luminosity therefore leads to a disk that is less susceptible to convective instability.
In the case of a solar mass star with a disk of $M_{\mathrm{disk}}=\SI{0.1}{\solmass}$ and $\alpha=10^{-3}$, convection zones span from $\sim \SIrange{0.3}{1.1}{\AU}$ and from $\sim \SIrange{3}{3.2}{\AU}$ close to the midplane. \\ \indent
An increase of the disk mass has the exact opposite effect on convectively unstable regions as can be seen in \autoref{VCI}b. As the mass increases, densities and therefore opacities increase, which heats up the disk at small radii. Again, the combination of stellar and accretion heating shifts the unstable zones to larger radii for larger disk masses, but due to the increase of densities at all radii, convection zones grow nonetheless.
We conclude that disks with higher total mass are more prone to be unstable due to convection.

As can be seen in \autoref{VCI}c, our disk model behaves quite similar when the disk's $\alpha$ parameter is increased. In that case, the efficiency of accretion heating becomes higher. Hence, temperature gradients increase and convection zones grow. As $\alpha$ becomes larger, the innermost parts of the disk become hotter and the sublimation lines move outwards. Convection zones grow as the disk is heated and temperature gradients become larger.
For an intermediate $\alpha$-value of $10^{-3}$, two spatially separated convection zones exist, which span from $\sim \SIrange{0.35}{1.1}{\AU}$ and from $\sim \SIrange{3}{4}{\AU}$ close to the midplane. 
It can be said that the more effective viscous heating becomes, the more susceptible a disk is to convective instability and thus indirectly to VSI.
The growth rates of the instability have been calculated by use of \autoref{vci_growth} and increase with height. They are generally in the order of  $\sim \SIrange{e-4}{1}{\om}$. \clearpage
\begin{figure}[ht] \centering
\gridline{
\fig{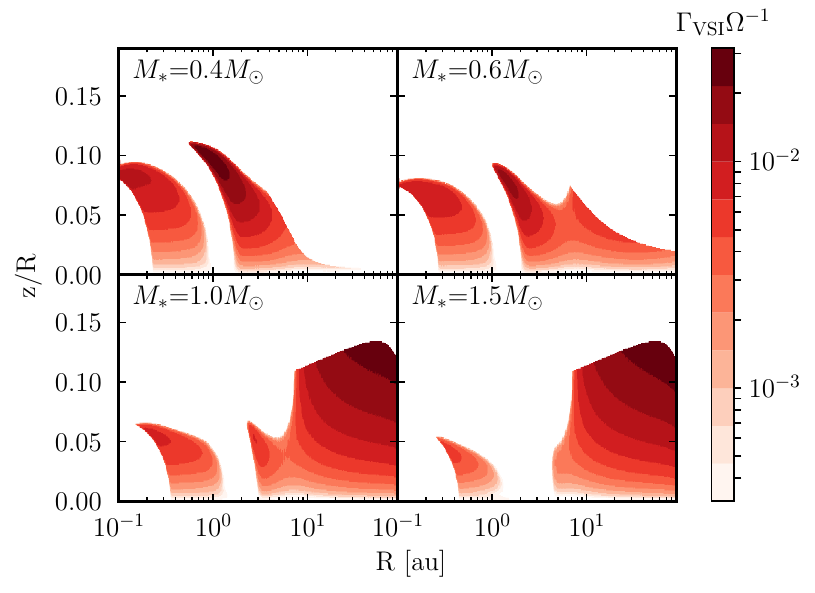}{0.455\textwidth}
{(a) VSI growth rates for different stellar masses}}
\gridline{
\fig{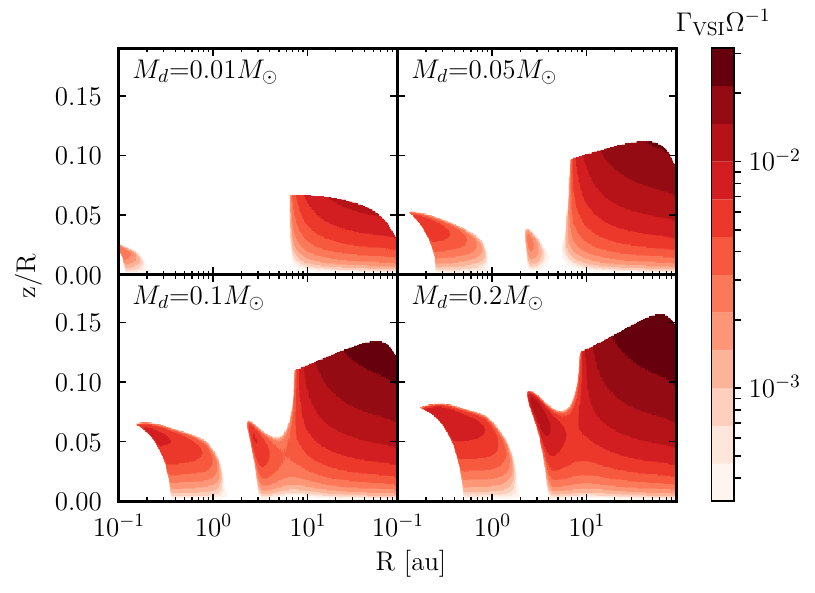}{0.455\textwidth}
{(b) VSI growth rates for different disk masses}}
\gridline{
\fig{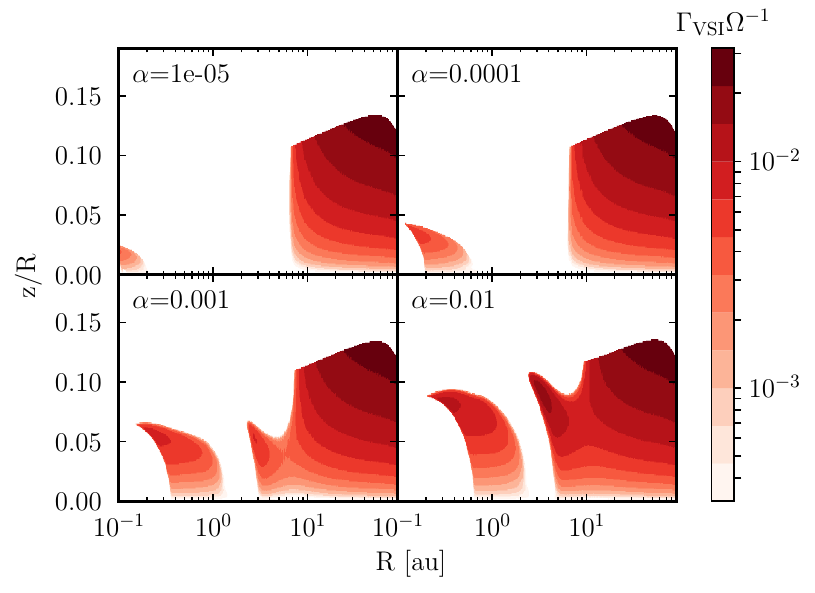}{0.455\textwidth}
{(c) VSI growth rates for different $\alpha$-parameters}
}
\caption{Growth rates of the VSI in units of the local dynamical time-scale ($\Gamma \Omega^{-1}$). One parameter is varied per panel, the other two are held constant at $M_*=\SI{1}{\solmass}$, $\alpha=10^{-3}$ or $M_{\mathrm{disk}}=\SI{0.1}{\solmass}$. No growth occurs in white areas \ADD{and in the midplane}.}
\label{VSI}
\end{figure}\clearpage

\subsection{Vertical Shear Instability}\label{subsec:vsi_res}
For a disk to be unstable due to VSI, a vertical gradient in angular velocity is required. The stratification has to be neutral in the sense of buoyancy, or at least allow for sufficiently fast thermal relaxation in order to overcome restoring buoyancy forces. A vertically buoyant structure also allows for the growth of VSI if $\tau_{\mathrm{relax}} \Omega\lesssim 10$ is fulfilled \citep{Nelson_2012}.
In order to calculate the growth rates for the VSI with \autoref{VSI_growth}, a certain radial wavenumber was chosen. We then used \autoref{VSI_max} to determine which corresponding vertical wavenumber leads to maximum growth of a perturbation. The results of \cite{Lin_2015} suggest, that modes with large \CORR{wavenumber} decay in viscous disks since their growth time becomes similar to the viscous time scale. They therefore draw the conclusion, that only perturbations with wavenumbers in the order of $k_R H \sim \mathcal{O}(10)$ will grow significantly, where $H$ refers to the disk's local pressure scale height. We chose $k_R H= 50$ for our investigation, because this value is $\sim \mathcal{O}(10)$ but also sufficiently large to allow for efficient cooling at larger radii (see \autoref{t_diff}). \ADD{Higher wavenumbers allow in principle for VSI in even larger areas beyond the adiabatic regions. A determination of the fastest growing wavenumbers, including the effects of realistic thermal relaxation models requires a numerical study like it was done by \cite{Lin_2015} and goes beyond the scope of this work. Our investigation of the VSI therefore relies on the arbitrary choice of a $k_R$ in the order of magnitude that was suggested by \cite{Lin_2015}.}
At larger distances to the star, the disk becomes optically thinner and stable to buoyancy, which renders it unstable to the VSI. 
As can be seen in \autoref{VSI}, growth rates increase with vertical distance to the midplane. Close to the midplane they reach values of $\sim$ \SIrange{e-4}{e-3}{\om} and growth rates of up to $\sim$ \SIrange{e-2}{e-1}{\om} at heights larger than $z/R \sim 0.05$.
For very faint, low mass stars, temperatures far away from the star are quite low ($T\lesssim 20 K$). It can be seen in \autoref{t_thin} that optically thin relaxation times are strongly temperature dependent ($\tau_{\mathrm{emit}}\propto  T^{-3}$). Therefore, optically thin cooling dominates the outer regions of the disk for low mass stars. This time scale can become so large, that criterion \eqref{VSI_crit} for VSI growth is no longer obeyed and no outer VSI-susceptible zone exists.
Temperatures rise when the solar mass is increased and the outer regions become susceptible for VSI, as can be seen in \autoref{VSI}a.
Therefore we find that the mass of the central star has different effects on the susceptibility to the VSI in the inner and in the outer parts of the disk. A low stellar mass (compared to the disk mass) leads to large convectively unstable zones, which are also VSI susceptible, but to slow thermal relaxation in the outer regions, which inhibits the growth there (\autoref{VSI}a upper panels).
Larger stellar masses lead to smaller convection zones but to fast thermal relaxation far away from the star, which has the opposite effect on the VSI. A generally hotter disk due to a more massive star therefore becomes susceptible to VSI at larger radii but less VSI active at smaller radii.

An increase of the disk's mass has the opposite effect on the inner VSI active regions. As described in the previous sections, low disk masses lead to a disk that is stable against buoyancy. This means that the inner VSI-susceptible regions are small for small disk masses and larger for large disk masses (see \autoref{VSI}b).
The outer susceptible region shows a different behavior. When the disk mass is very small ($M_{\mathrm{disk}}\sim \SI{0.01}{\solmass}$), densities also become small at larger radii. Consequently, collisions between molecules and dust particles become rare, and thermal relaxation slows down.
This is why the outer VSI-susceptible region is smaller for smaller disk masses.

The $\alpha$ parameter defines the efficiency of viscous heating in the disk. Increasing this parameter therefore leads to enlarged convectively unstable zones, with the same effect on VSI active zones (see \autoref{VSI}c). The shape and size of the outer susceptible region depends mostly on thermal relaxation and therefore remains mostly unaffected by a variation of $\alpha$.

All in all, VSI active regions close to the star coincide with vertical convection zones and have the same dependency on the system parameters as convection. The extent of the outer susceptible zone depends on cooling times and is therefore favored in systems with high stellar mass, disk mass and $\alpha$-parameter.

\subsection{Stability Maps}\label{subsec:stability}
\begin{figure}[ht] \centering
\gridline{
\fig{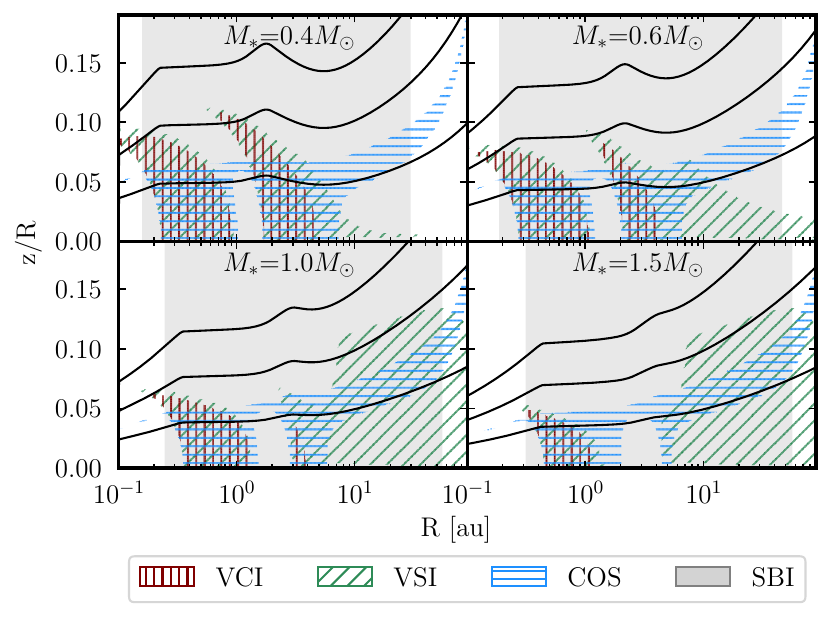}{0.455\textwidth}
{(a) Stability Maps for different stellar masses}}
\gridline{
\fig{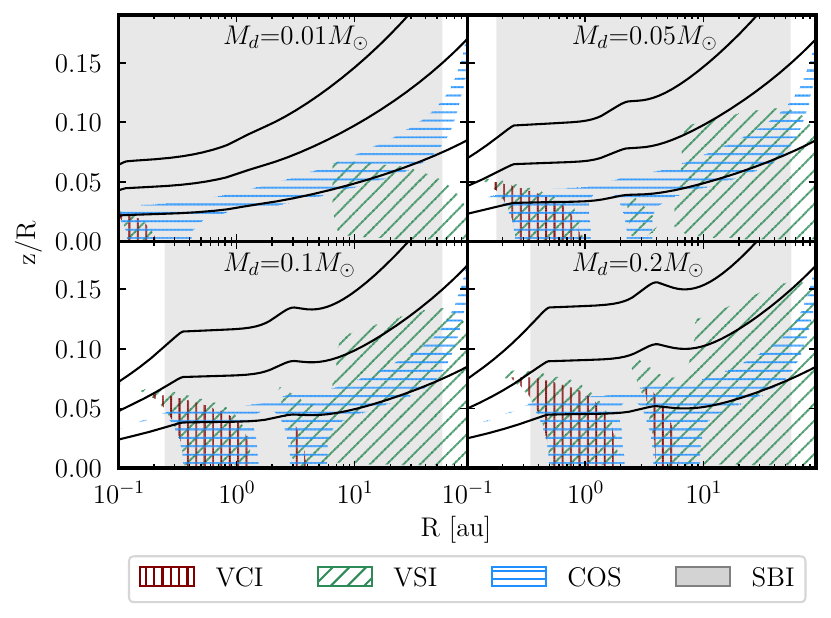}{0.455\textwidth}
{(b) Stability Maps for different disk masses}}
\gridline{
\fig{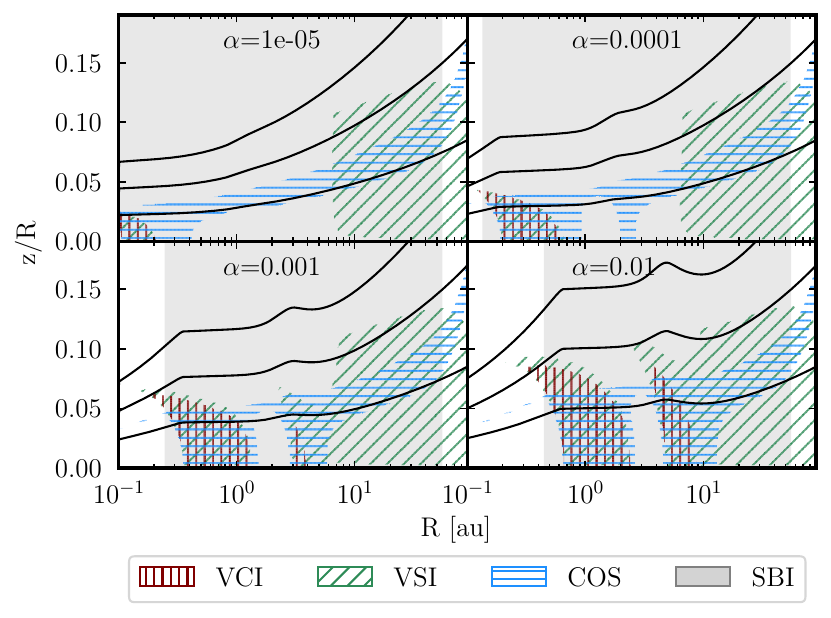}{0.455\textwidth}
{(c) Stability Maps for different $\alpha$-parameters}
}
\caption{Stability Maps for the models presented in this work. One parameter is varied per panel, the other two are held constant at $M_*=\SI{1}{\solmass}$, $\alpha=10^{-3}$ or $M_{\mathrm{disk}}=\SI{0.1}{\solmass}$. The black lines represent 1, 2 and 3 pressure scale heights.}
\label{Maps}
\end{figure}

The spatial distributions of the four discussed instabilities and the investigated parameter sets are summarised in \autoref{Maps}.
It can be seen that the COS and the VCI share large parts of their susceptible regions, since both of them rely on the opacity structure of the disk in similar ways. 
The VSI operates also in the convectively unstable region, since vertical perturbations are enhanced there, as well as at radii larger than $\sim \SI{10}{\AU}$ where the disk becomes optically thinner.
\ADD{All hydrodynamical instabilities are favored in set-ups with small $M_*$ and large $\alpha$ and $M_{\mathrm{disk}}$. These are the disks in which the temperature gradients are dominated by viscous heating instead of stellar irradiation. A disk that is viscously heated thus becomes susceptible to hydrodynamic turbulence in large parts. Disks with very low $\alpha \lesssim 10^{-4}$ or $M_{\mathrm{disk}}$ are nonetheless unstable to VSI at larger radii and can develop COS in a thin region at $z \gtrsim \SI{1}{H}$. If these instabilities are able to produce finite amplitude perturbations, they might be able to develop into large scale SBI vortices, which are amplified in large parts of the disk.}

\section{Summary \& Conclusions}\label{sec:conclusion}
In this paper, we investigate the stability of active protoplanetary disks (where the temperature profile is set by accretion) as well as passive disks (in which the  temperature profile is dominated by irradiation) by use of 1+1D steady state accretion disk models including stellar irradiation. This allows for the treatment of flux-limited radiative transfer, caused by viscous heating and makes it possible to apply detailed models of the local rate of thermal relaxation, thus making it possible for the first time to the authors' knowledge, to spatially map the growth rates of the COS, the SBI and the VSI to the radial-vertical plane of a realistically stratified circumstellar disk.

We found that we can reproduce the radial temperature profile in full time-dependent 3D axissymmetric radiation hydro simulations as performed by \cite{Bitsch2015} for their model parameters, but can cover a much wider parameter space.

It is shown that even an almost quiescent disk, with an extremely low viscosity parameter of $\alpha=10^{-5}$ becomes unstable to COS and VSI at radii $\gtrsim \SI{10}{\AU}$ and for the COS even closer to the star at heights of $\sim \SI{1}{\scaleheight}$ above the midplane. The turbulent structures, resulting of such instabilities were shown to grow to large scale vortices by various authors \citep{Meheut_2012, Lyra_2014, Manger_2018}.

A closer look was therefore also taken upon the growth rates and susceptible regions of these structures by consideration of the SBI mechanism. Our results show, that the vertically integrated, radial stratification of the disk allows for positive growth rates over almost the whole radial extent of the disk, between $\sim \SIrange{0.3}{50}{\AU}$ at timescales of $\sim \SI{1000}{\orb}$ even if $\alpha$ has low values of $10^{-5}-10^{-4}$. Vortices which evolve at large radii, due to the perturbations caused by COS and VSI might thus be able to migrate towards the central star while being constantly forced by the SBI contributing to their longevity \citep{Lesur_Papaloizou_2010, Paardekooper_2010}. These findings also indicate that the SBI-mechanism could be a controlling mechanism for the frequently observed vortices in protoplanetary disks \citep{Carrasco_2016,Marel_2013}.  

Inner regions of disks which reach $\alpha$-values of $10^{-4}$, are already subject of strong enough viscous heating and evolve into a radially and vertically buoyant structure, as was shown in  \autoref{subsec:cos} and \autoref{subsec:vci}. 
We have shown that these buoyantly unstable zones arise wherever the disk becomes optically thick enough to allow for the existence of strong vertical temperature gradients which means that they are larger in disks with a high mass ($\gtrsim \SI{0.01}{\solmass}$ for a $\SI{1}{\solmass}$ star). 

The resulting radial buoyancy, in combination with thermal relaxation is able to drive COS at maximal growth rates of $\sim \SI{e-3}{\om}$. COS is generally favored by massive disks around low mass stars and high $\alpha$-viscosity.
\CORR{The opacity structure and thus the relaxation time criterion determine the fastest growing modes, which means that short wavelength perturbations (large $k$) grow best in optically thick regions and long wavelength perturbations grow better in optically thinner regions.}

VSI depends even stronger on radiative cooling and thus requires the optically thick parts of the disk to be at least buoyantly neutral, which impedes repelling forces on vertical perturbations. The internal heat production, caused by $\alpha$ values of $10^{-4}$, is sufficient to provide such a disk structure in the denser interior zones and therefore makes it possible for the VSI to arise, even if the necessary criterion for cooling by \cite{Lin_2015} is not fulfilled.
The finding that VSI can produce an $\alpha = 10^{-4}-10^{-3}$ \citep{Nelson_2012, Stoll_2014,Manger_2018} enables the possibility that the VSI can maintain the thermal structure of the disk that it needs to operate. What needs to be shown, is at what height in the disk, the thermal energy will actually be released.

Two distinct regions of VSI growth exist at small radii between $\sim \SIrange{0.3}{1.1}{\AU}$ and $\sim \SIrange{3}{3.2}{\AU}$ (for model parameters of $M_{*}=\SI{1}{\solmass}$, $M_{\mathrm{disk}}=\SI{0.1}{\solmass}$, and $\alpha=10^{-3}$), which coincide with the regions unstable to vertical buoyancy. Larger disk masses and $\alpha$-parameters enhance viscous heating and therefore also increase the spatial extents of these inner susceptible zones.
Fast cooling is required to allow for VSI far away from the central object, where the vertical stratification is stabilised by weak stellar irradiation and the absence of viscous heating. Thermal relaxation in these parts of the disk is limited by either collisional or radiative timescales. We have shown, that low mass disks ($M\lesssim \SI{0.01}{\solmass}$) have too small gas densities at radii $\gtrsim \SI{10}{\AU}$ to allow for frequent enough collisions between emitters and carriers of thermal energy to cool the disk sufficiently fast. The VSI is therefore strongly hampered in such setups.
The gas far away from low mass stars also has low rates of thermal relaxation, since the low temperatures of gas and dust particles do not allow for efficient emittance of energy via black body radiation. 
We have therefore shown that VSI at large distances from the star is strongly present for large $M_{*}$ and $M_{\mathrm{disk}}$ whereas its presence at smaller distances is determined by the vertically polytropic structure, and therefore favored by small $M_{*}$, large $\alpha$ and $M_{\mathrm{disk}}$.
The growth rate of the instability increases vertically from $\sim \SIrange{e-4}{e-2}{\om}$.

Future work should deal with more recent opacity models, including the evolution of the dust component of the disk, and a variable chemical composition of the disk, having influence on the adiabatic coefficient $\gamma$. Especially the latter one probably has an important influence on the strength of buoyancy driven instabilities. Then it also should be possible to map out possible Zombie Vortex regions \citet{2018arXiv181006588B}. 

Passive disks with an assumed surface density profile inspired by the so called minimum mass solar nebula \citep{Chiang1997}, i.e. $\beta_\Sigma = -1.5$ and radial temperature gradient determined only by irradiation will have a radial increasing entropy structure and therefore not be the subject of SBI. But note that this steep density profile was extrapolated from the  ``solid'' mass distribution in our solar system and not from the gas distribution around the young sun, which should be completely different in the course of dust growth and pebble drift \citep{Birnstiel2012}. Modeling of an actively accreting disk, be it as simple as in our model, or a more elaborate 1+1D irradiation model as in \cite{Dalessio2006} or even in a full 2D hydrodynamic model gives a much shallower surface density profile than $\beta_\Sigma = -1.5$, in range of the values derived from observations \citep{Andrews2010} of $\beta_\Sigma = -1.1$ to $-0.4$. These values predict a wide region in a disk to have a negative, buoyantly unstable gradient in vertically integrated entropy, i.e. the condition for convective amplification of vortices (SBI). \\ \indent
Our model is in general agreement with the more complicated models of disk structure \citep{Dalessio2006,Bitsch2015} if one considers about the first two pressure scale heights above the midplane. 
Our predictions for stability/instability above that region is less reliable. Here we suggest additional work to investigate cooling rates and entropy structures in these dilute regions, but remember that those regions might also be influenced by magnetic fields because of their sufficient ionisation state \cite{Dzyurkevich_2013}, yet being hampered by ambipolar diffusion. Typically those would be the regions where a wind is being launched from the disk in an interplay of photo-evaporation and magnetic fields \citep{Pudritz_2007,Koenigl_2010}. 

We conclude that hydrodynamical instabilities can exist in large portions of protoplanetary disks and that they benefit from the release of accretion power if at least a fraction of the released heat gets deposited within 1 or 2 pressure scale heights around the midplane. But note that even disks with very low accretion rates have a radial temperature stratification, which renders them unstable to SBI and partially unstable to VSI and COS, especially at radii $\gtrsim \SI{10}{\AU}$.
Observed disk profiles by \citet{Andrews2010} are indeed unstable to the SBI.

If the resulting $\alpha$-values from either these hydro instabilities or non-ideal MHD effects, are low to moderate ($\alpha=10^{-4}-10^{-3}$), as suggested by \cite{2013ApJ...765..115R}
and the heat is deposited around the midplane, then also regions closer to the star become vertically and radially buoyant and therefore susceptible for COS and VSI. 

Thus, the largest caveat in our work is, that even if sufficient $\alpha$ values are measured in simulations of SBI and VSI \citep{2013ApJ...765..115R, Nelson_2012,Stoll_2014} it is currently not known where the kinetic energy resulting from the release of potential energy in the accretion disk is deposited. If at least a part of this energy is deposited close to the midplane of the disk, then hydrodynamical turbulence has a good chance to operate in large parts of the planet forming regions of protoplanetary disks, even in the midplane, where non-ideal MHD effects damp otherwise dominant magnetic effects sufficiently \citep{LyraKlahr_2012}. With its well known properties of forming vortices \citep{2013ApJ...765..115R,Manger_2018} and zonal flows, hydrodynamic instability can be a major agent in forming planetesimals and thus determine the properties of planetary systems \citep{Klahr2018}.

\section*{Acknowledgements}
We would like to thank the whole planet and star formation theory group of the Max-Planck-Institute for Astronomy in Heidelberg for many fruitful discussions of the topic, their help and advice. We especially thank Natascha Manger, Martin Schlecker, Bertram Bitsch and Henrik Latter for useful suggestions, discussions and advice.

\begin{figure*}[ht]
\centering APPENDIX
\includegraphics[width=\textwidth]{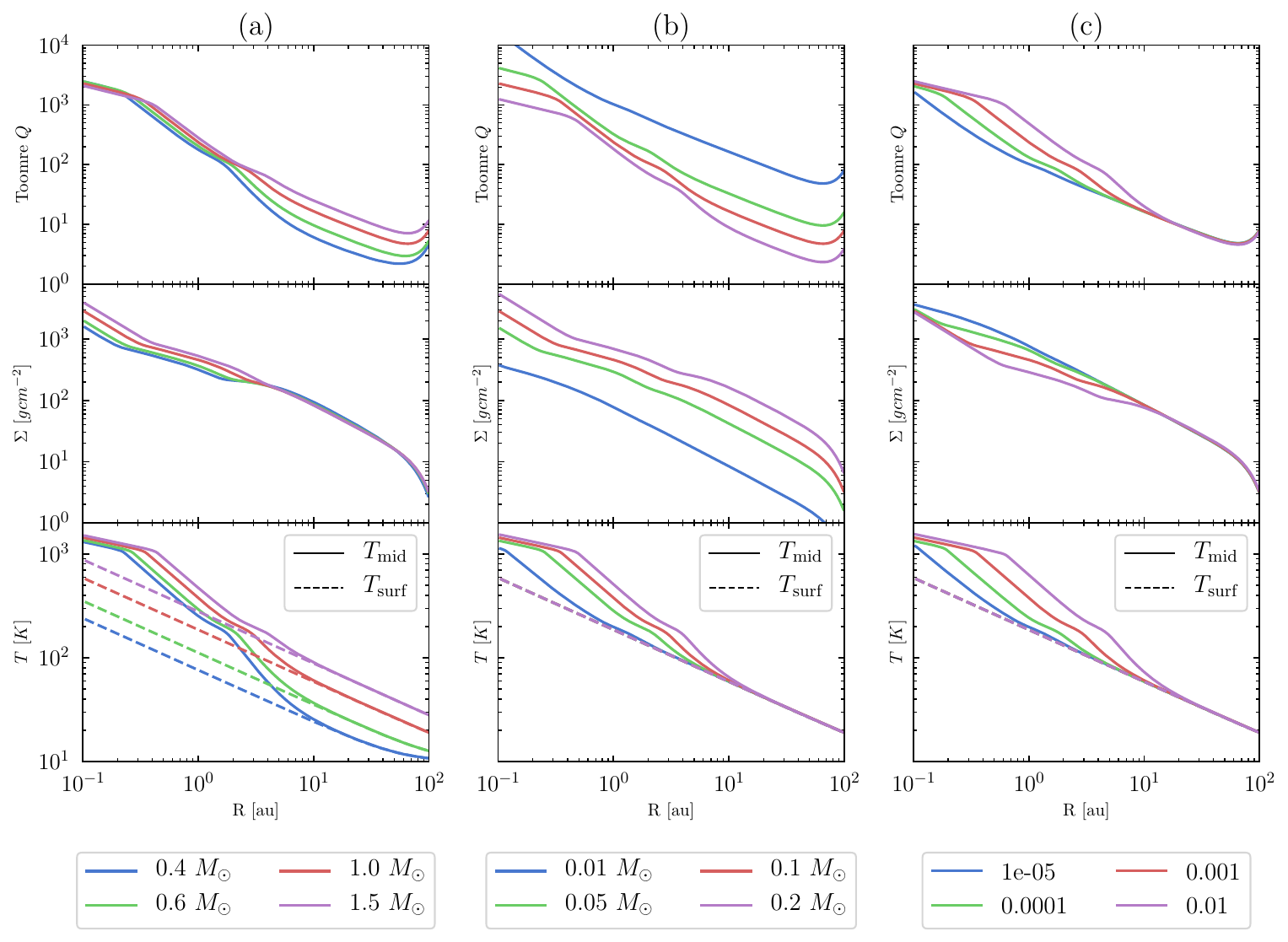}
\caption{Toomre $Q$ (top row), column density $\Sigma$ (middle row), midplane temperature $T_{\mathrm{mid}}$ and surface temperature $T_{\mathrm{surf}}$ (bottom row) as a function of radius for the presented disk models for the different stellar parameters in column (a), disk masses in column (b) and $\alpha$-paramters in column (c). It can be seen that every model is stable in the sense of $Q>1$. Increasing disk temperatures lead to higher $Q$ values for higher stellar mass or $\alpha$-parameter. Larger disk masses and thus column densities reduce $Q$. We also added the irradiation temperature $T$ to the bottom plots. This enables us to identify the transition radius from irradiation dominated disks (passive) to accretion heating dominated disks (active), which is a function of disk mass, accretion rate ($\dot{M} \propto \alpha M_{\mathrm{disk}}$) and opacities.} 
\label{Toomre}
\end{figure*}

\begin{figure*}[ht] \centering
\gridline{
\fig{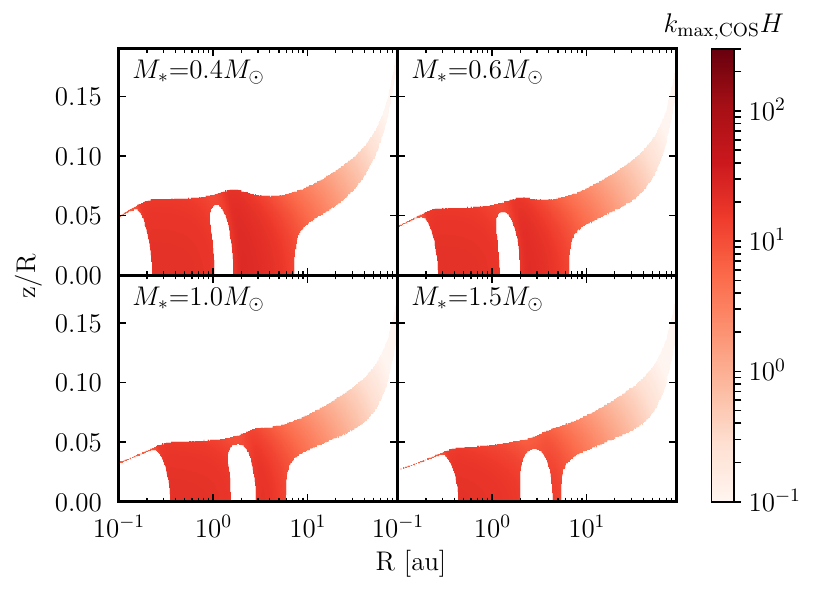}{0.455\textwidth}
{(a) max. growing COS wavenumbers for different stellar masses}}
\gridline{
\fig{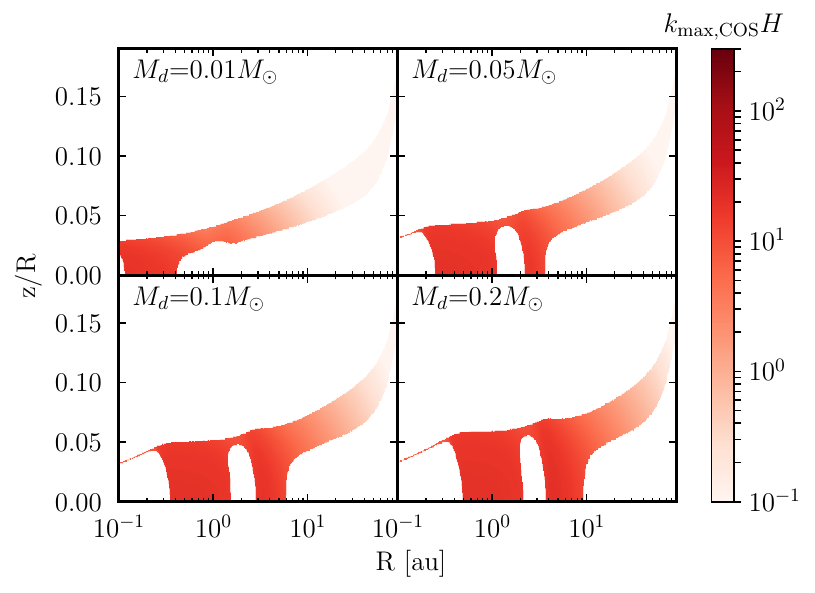}{0.455\textwidth}
{(b) max. growing COS wavenumbers for different disk masses}}
\gridline{
\fig{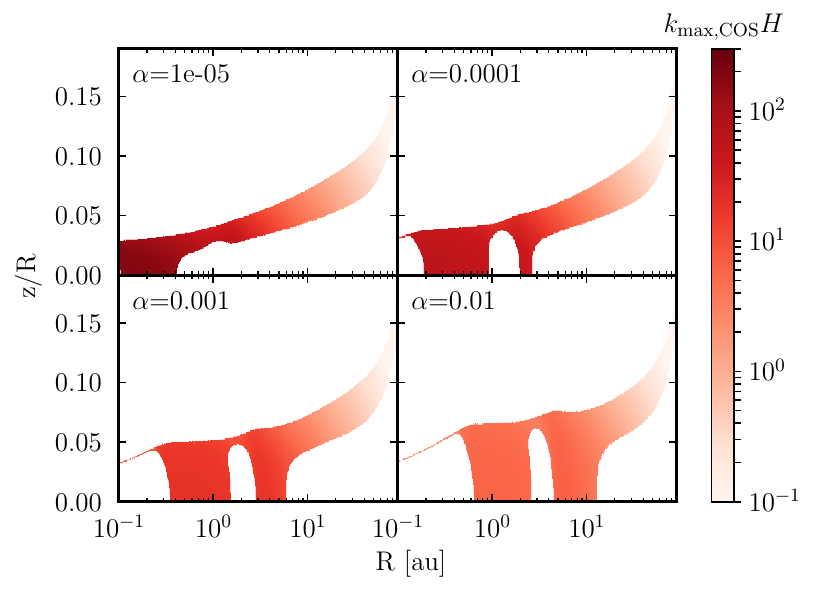}{0.455\textwidth}
{(c) max. growing COS wavenumbers for different $\alpha$-parameters}}
\caption{Maximally growing wavenumbers of the linear COS, corresponding to \autoref{COS}. One parameter is varied per panel, the other two are held constant at $M_*=\SI{1}{\solmass}$, $\alpha=10^{-3}$ or $M_{\mathrm{disk}}=\SI{0.1}{\solmass}$. No growth occurs in the white areas.}
\label{COS_k}
\end{figure*}

\begin{figure*}
\centering
\includegraphics[width=\textwidth]{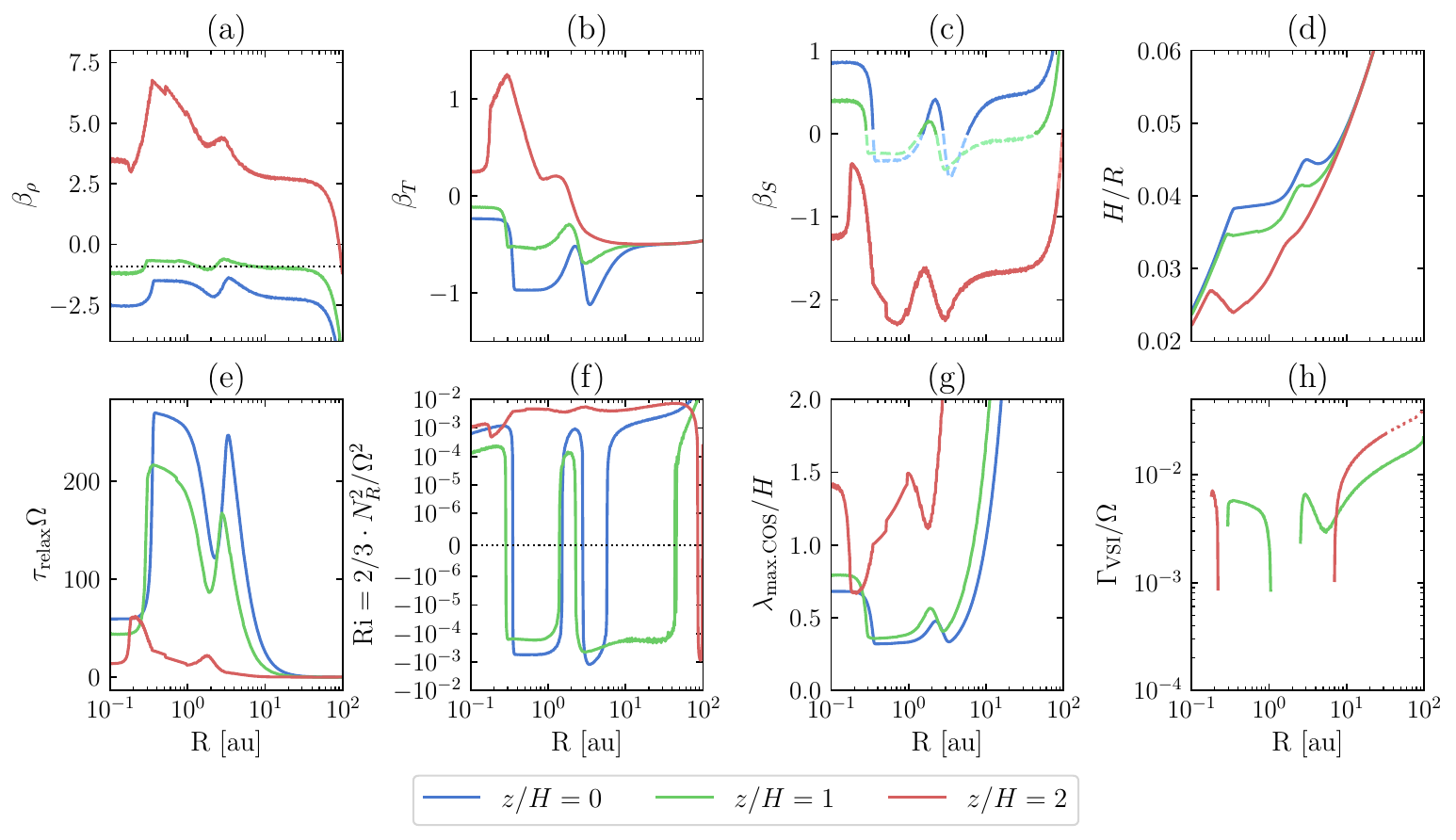}
\caption{Several Disk properties of a model calculated with a disk mass of $M_{\mathrm{disk}}=\SI{0.1}{\solmass}$, stellar mass of $M_*=\SI{1}{\solmass}$ and $\alpha=10^{-3}$ for the midplane and at $z=H$ and $z=2H$. \\
Panel (a) shows the logarithmic radial density gradient, panel (b) displays the radial logarithmic temperature gradient. \\
(c) shows the radial entropy gradient. Here it can be seen that entropy gradients are completely negative at $z=2H$ above the midplane. However,  pressure gradients become radially positive, which means that the disk is radially stable against convection at $z=2H$, in the sense of the Schwarzschild criterion. At a height of $z\sim 1H$ and in the midplane, pressure and entropy gradients are partially parallel (shown as the dashed section in the green line). This means, that these regions are unstable in the sense of the Schwarzschild criterion, which can give rise to COS. \\
Panel (d) shows the local scale height profile. \\
In Panel (e) cooling times for $k=1/H$ are shown. It can be seen that thermal relaxation becomes faster with height above the midplane, since densities decrease. The regime of collisionally limited relaxation due to extremely low densities is not covered at these heights and becomes dominant at $z\sim 3 H$. \\
Panel (f) shows the radial Richardson number of the disk. The convectively unstable zone at $z= 1H$ is visible, since negative Richardson numbers indicate instability in the sense of the classical Schwarzschild criterion without rotation taken into account. A Richardson number larger than $-\nicefrac{2}{3}$ indicates radial stability in the sense of the standard Solberg-H\o iland criteria ($N_R^2+\Omega^2>0 \quad\Rightarrow$ stability).\\
Panel (g) shows the perturbation wavelength for which of the COS' growthrate becomes maximal: \\
$\gamma \tau_{\mathrm{relax}}\Omega =1 \quad \Rightarrow \quad \lambda_{\mathrm{max. COS}}/H=\sqrt{4  \pi^2 \tilde{D}_{\mathrm{E}}/(\gamma \Omega H^2)}$ with $\tau_{\mathrm{relax}}=1/(k^2 \tilde{D}_{\mathrm{E}})$. \\
In panel (h) VSI growth rates are displayed. The dotted parts indicate that growth is inhibited by cooling times $> \tau_{\mathrm{crit}}$ \citep{Lin_2015}}
\end{figure*}

\clearpage
\bibliography{Literatur}

\end{document}